\documentclass[11pt,twoside]{article}
\usepackage{latexsym}
\usepackage[english]{babel}
\usepackage{amssymb,amsfonts,amsmath}
\usepackage{mathrsfs}
\usepackage{fancyhdr}
\usepackage{feynmp}
\usepackage{graphicx}

\newtheorem{theorem}{Theorem}[section]

\newtheorem{lemma}[section]{Lemma}
\newtheorem{definition}[section]{Definition}
\newtheorem{proposition}[section]{Proposition}

\begin{document}

\title{Weak-Coupling Limit. I \\ A Contraction Semigroup for Infinite Subsystems}

\author{David Taj\\ {\small Dept. Physics, Politecnico di Torino, C.so Duca degli
Abruzzi 24, 10129, Torino, Italy} \\ david.taj@gmail.com}

\maketitle

\begin{abstract}
We consider the class of quantum mechanical master equations defined on a generic Banach space, arising by projecting weakly perturbed one-parameter groups of isometries. We show that the possible semigroup approximations are far from unique. However, uniqueness can be reestablished through the introduction of a dynamical time averaging map. The generator of the resulting Contraction Semigroup is always well defined, irrespective of the dimensions of the projected subspace, and of the spectral properties of its free dynamics. We show how our approach includes and generalizes the preexisting literature.
\end{abstract}

\section{Introduction}
After the pioneering works in 1974 and 1976 by
Davies~(\cite{davies1,davies2}), a huge amount of physical
information about the irreversibility and the evolution of open
quantum mechanical systems has been gained~\cite{spohn}. Alicky~\cite{alicki}
showed in 1977 that these efforts were deeply connected to the
celebrated "Fermi's Golden Rule", that now had become
mathematically consistent. The conceptual importance of these
works is clearly not only academic, as the need for a better
understanding of irreversible processes has never been more
urgent. Today, so many nanotechnologies are pushing devices
towards limits where neither quantum phase coherence, nor
dissipation/dephasing, can be neglected~\cite{SL,QD1,QD2,QD3,RMP}. Many
attempts to improve the theory have been made since then (see for
example~\cite{breuer,open}), but despite the compelling need, no
substantial, fundamental progress, directly applicable to nowadays
technologies, has been made so far.

To be more specific, the problem is to understand the dynamics of
a subsystem of interest, when the global system is fully coherent.
In many cases, the dynamics of the global system can be splitted
into a part that leaves the subsystem invariant, plus an
interaction between the subsystem and the remaining
"unobserved" degrees of freedom. The problem then arises
whether or not the subsystem can be given a markovian, possibly
dissipative, dynamics as a consequence of its interaction with
those degrees of freedom. It has been shown since the '70s that this problem can be attacked, and partially solved~\cite{davies1,davies2}, when the amount of unobserved degrees of freedom is huge,
and the interaction is made small. This last condition is referred
to as the "weak-coupling limit".

In~\cite{davies1} the author was able to give a solid physical
model of a discrete "atom" (system A) interacting with a fermionic
particle reservoir at thermal equilibrium (environment B). In that
case, the subsystem was made of unentangled pairs of atomic states
(also referred to as "density matrices") and a fermionic thermal
equilibrium state, while all the entangled pairs just constituted
the remaining, uninteresting, degrees of freedom. The model was of
high conceptual importance, as the markovian (and dissipative)
dynamics for the subsystem was shown to guarantee the state
positivity at all times. This fact gave just enough internal
robustness to the model as to be of invaluable practical use, for
at any time, the subsystem evolution could be given a strong
physical meaning. But the system A had to be finite dimensional,
or at least, its unperturbed hamiltonian was forced to have
discrete spectrum. This fact constitutes a severe limitation with respect to nowadays needs to explore systems at mesoscopic scale, where energy spectra are often of mixed nature.

Unfortunately, positivity through the weak-coupling limit procedure becomes no more available when the
system is infinite dimensional, or the spectrum is continuous. In that case indeed Davies showed
that a markovian approximation, for the subsystem exact projected
dynamics, could still be achieved \cite{davies2}, but positivity
could not be shown even
for the old and "safe" partial tracing over the thermal bath's
degrees of freedom. Being more general, the theory was and is currently applied in many physical urgent contests (see for example~\cite{rossi}), but certainly it did not
share the enormous success of the previous one among physicists,
precisely because of the serious positivity limitation. For example, all of
the steady state analysis became, physically, completely
meaningless.

In this work we provide a rigorous solution to this problem in the general contest of Banach spaces, by extending and putting on more firm grounds our initial study in~\cite{tajpra}. We suppose the
global system is undergoing a fully coherent evolution (according to a perturbed one-parameter group of isometries), and discover a whole class of new markovian approximations for the subsystem, by suitably manipulating the memory kernel of the exact projected evolution, in the limit of small perturbation. This class includes the markovian approximation performed by Davies in~\cite{davies2}, and every generator in it is well defined irrespective of the subsystem spectral properties and dimensions. Since we will work in the general contest of Banach spaces, we will not address positivity at this time, but it will be evident that none of generators found so far gives rise to a contraction semigroup, which is a basic requirement for a Quantum Dynamical Semigroup~\cite{lindblad} in the more specific contest of $W^*$-algebras.

At this point, we will introduce a "transition time" function, that will scale with the inverse of the perturbation, and will represent physically the transition time among the set of relevant states, due to the perturbation. This time will serve us to perform a "dynamical average" among all the generators previously found. The result will be a "dynamical time averaging map", very similar to the time average proposed in~\cite{davies1}, but fundamentally different in that $i)$ it will scale with the coupling constant and $ii)$ it averages among different generators (in~\cite{davies1} the time average acts on one generator only). We shall be able to prove that the resulting transition time dynamically averaged generator $i)$ correctly approximates the exact projected dynamics in the weak-coupling limit, $ii)$ is always well defined, irrespective of the subsystem spectral properties and dimensions, $iii)$ accounts also for first order contributions, $iv)$ boils down to the averaged generator in~\cite{davies1} in case of discrete spectrum, and we shall prove the all important result that $v)$ it gives rise to a contraction semigroup on the projected Banach subspace.

Thus the requirement of a contraction semigroup essentially removes the degeneracy of all the possible semigroup approximations. The generality with which we shall be able to make our statements (no dependence on spectral properties or dimensions of the projected subspace) establishes on firm grounds the possibility of many new important physical (and mathematical) applications.

\section{General Theory and Motivation}
We report from Davies \cite{davies2} the general framework we'll
be involved with. We suppose that $P_0$ is a linear projection on
a Banach space $\mathcal{B}$ (that represents some global system),
put $P_1=1-P_0$ and $\mathcal{B}_i=P_i\mathcal{B}$, so that
\begin{equation}
\mathcal{B}=\mathcal{B}_0\oplus\mathcal{B}_1,
\end{equation}
and we take $\mathcal{B}_0$ to represent a subsystem of interest,
$\mathcal{B}_1$ being the remaining degrees of freedom. We suppose
that $Z$ is the (densely defined) generator of a strongly
continuous one-parameter group of isometries $U_t$ on
$\mathcal{B}$ with
\begin{equation}
U_t P_0=P_0 U_t
\end{equation}
for all $t\in\mathbb{R}$, or equivalently
\begin{equation}
[Z,P_0]=0
\end{equation}
and put $Z_i=P_i Z$. We suppose that $A$ is a bounded perturbation
of $Z$ and put $A_{ij} = P_iAP_j$. We let $U^\lambda_t$ be the one
parameter group generated by $(Z + \lambda A_{00} + \lambda
A_{11})$ so that
\begin{equation}
U^\lambda_t P_0=P_0 U^\lambda_t
\end{equation}
for all $t\in\mathbb{R}$, and let $V_t^\lambda$ be the one
parameter group generated by $Z+\lambda A$, so that
\begin{equation}\label{eq:iteration_semigroup}
V^\lambda_t = U^\lambda_t + \lambda \int_{0}^t U^\lambda_{t-s}
(A_{01}+A_{10})V^\lambda_s ds.
\end{equation}
Then putting
\begin{equation}
X^\lambda_t=P_0 U^\lambda_t
\end{equation}
and defining the projected evolution as
\begin{equation}
W^\lambda_t=P_0 V^\lambda_t P_0
\end{equation}
and one obtains the all important closed and exact integral
equation
\begin{equation}\label{eq:exact}
W^\lambda_t=X^\lambda_t+\lambda^2 \int_{0}^t ds \int_{0}^s du\;
X^\lambda_{t-s}A_{01}U^\lambda_{s-u} A_{10} W^\lambda_u .
\end{equation}
This is nothing but the integrated form of the well known master
equation constructed by Nakajima, Prigogine, Resibois, and Zwanzig
\cite{naka,zwanzig}, which follows by repeatedly making use of
(\ref{eq:iteration_semigroup}) with
\begin{equation}
W^\lambda_t = X^\lambda_t + \lambda \int_{0}^t ds\;
X^\lambda_{t-s} A_{01}V^\lambda_s P_0
\end{equation}
and
\begin{equation}
P_1 V^\lambda_s P_0 = \lambda \int_{0}^s du\; U^\lambda_{s-u}
A_{10}W^\lambda_u.
\end{equation}
Now let $X^\lambda_t$ is a one parameter group of isometries. For example, Davies proves \cite{davies2} that
\begin{lemma}\label{lemma:davies}
If $\|P_0\|=1$ and $\|e^{At}\|=1$ for all $t\in\mathbb{R}$, then
$X^\lambda_t$ is a one parameter group of isometries on
$\mathcal{B}_0$ for all real $\lambda$.
\end{lemma}

Then, changing variables to $x=s-u$, $\sigma=\lambda^2 u$ and
introducing the time rescaled (and $A_{00}$-renormalized)
interaction picture evolution
$W^{\lambda,i}_\tau=X_{-\lambda^{-2}\tau}
W^\lambda_{\lambda^{-2}\tau}$, one is led to \cite{davies2}:
\begin{equation}\label{eq:davies_int}
W^{\lambda,i}_\tau=1+ \int_{0}^\tau d\sigma\;
X_{-\lambda^{-2}\sigma}^\lambda
K(\lambda,\tau-\sigma)X_{\lambda^{-2}\sigma}^\lambda
W^{\lambda,i}_\sigma
\end{equation}
where
\begin{equation}
K(\lambda,\tau)=\int_0^{\lambda^{-2}\tau} X_{-x}A_{01}U^\lambda_x
A_{10}\; dx.
\end{equation}
This form separates an "interacting" and "slowly varying" part
$K(\lambda,\tau)$ from the "rapidly oscillating" free-evolution
$X_{\lambda^{-2}\sigma}$ to second order in the coupling constant
$\lambda$. Now in the weak-coupling limit $\lambda\rightarrow 0$,
the slowly varying integral kernel $K(\lambda,\tau)$ converges to
\begin{equation}\label{def:K_D}
K_D=\int_0^{\infty} U_{-x}A_{01}U_x A_{10}\; dx \;,
\end{equation}
where $K_D$ is the celebrated Davies' superoperator. Substituting
$K_D\sim K(\lambda,\tau)$ in (\ref{eq:davies_int}) and moving back
to the "Schr\"odinger picture" we obtain the markovian
approximation for our subsystem dynamics
\begin{equation}
{W}^\lambda_{t}\approx \overline{W}^\lambda_{t}=e^{(Z_0+\lambda
A_{00}+\lambda^2 K_D)t}, \quad 0\leq t \leq \lambda^{-2}\tau \quad,
\lambda \approx 0
\end{equation}
Indeed in \cite{davies2} an important theorem shows that under
reasonable and general conditions the approximation holds in the
weak-coupling limit, up to $\lambda^{-2}$-rescaled positive times $\tau$,
independently of the subsystem dimensions or spectral properties.
Unfortunately, $K_D$ does not guarantee positivity of the
generated evolution, for example in the case of partial tracing
over a bath \cite{davies1}, and thus may fail to furnish a
physically acceptable model at times $t>\lambda^{-2}\tau$.
This fact precludes the important possibility to study the limit
dynamics at all times, together with all the steady state
analysis.

However, in \cite{davies1} the author shows that, in case
$\mathcal{B}_0$ is finite dimensional (or more generally when
$Z_0$ has discrete spectrum), one can define a temporal average
$K_D^\natural$ of $K_D$ and show that the generated evolution is
still asymptotic to the exact one in the weak-limit (at least when
$A_{00}=0$). Moreover, the author also shows that if the
projection $P_0$ is taken to be the partial trace over a bath, the
averaged $K_D^\natural$ generates a Quantum Dynamical Semigroup
(QDS) in the sense of \cite{lindblad}.

So finally $K_D$ does not guarantee a positive evolution at all
times, while $K_D^\natural$ is well defined only in the far too
restrictive case of discrete $Z_0$-spectrum. This serious problem
is physically so important as to motivate our work.

\section{Classes of Generators and Dynamical Transition Time}
\begin{figure}[htb]
\includegraphics[width=9cm,angle=0]{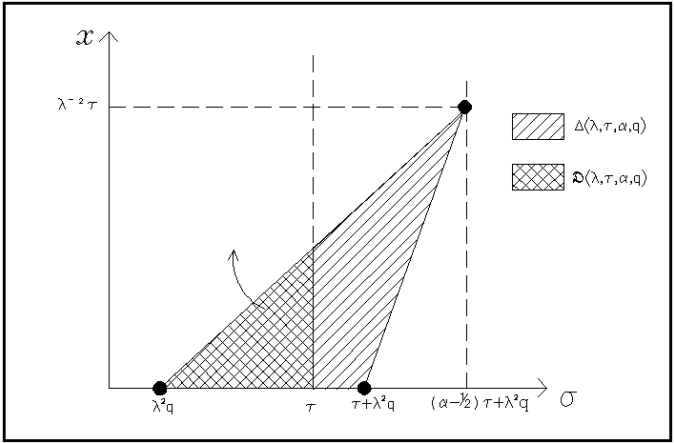}
\caption{Integration domains $\mathcal{D}(\lambda, \tau, \alpha,
q)$ and $\overline{\mathcal{D}}(\lambda, \tau, \alpha, q)$ defined in (\ref{def:domain}) and (\ref{def:domain_truncated}) respectively (we have put $q>0$ , $\alpha<-1/2$ for clarity). The arrow indicates the asymptotic behavior of the two domains in the weak coupling limit $\lambda\rightarrow 0$ (see discussion in the text following (\ref{def:domain})). \label{domain1}}
\end{figure}

In our first main theorem, we shall look for the most general form
of an operator whose evolution is still asymptotic to the exact
one in the weak-limit. To this purpose, note that $K_D$ has been
defined thanks to the change of variable
\begin{equation}\label{eq:vartr_davies}
\left(
\begin{array}{c}
s \\
u
\end{array}
\right) = \left(
\begin{array}{c c}
\lambda^{-2} & 1 \\
\lambda^{-2} & 0
\end{array}
\right) \left(
\begin{array}{c}
\sigma \\
x
\end{array}
\right).
\end{equation}
Here instead we would like to allow for the most general change of
variable that keeps a $\lambda^{-2}$ jacobian, proper of a second
order approximation, while fixing the relative variable to be
$s-u=x$. So we chose
\begin{equation}\label{eq:vartr_RT}
\left(
\begin{array}{c}
s \\
u
\end{array}
\right) = \left(
\begin{array}{ccc}
\lambda^{-2} & &\alpha+{1\over 2} \\
\lambda^{-2} & &\alpha-{1\over 2}
\end{array}
\right) \left(
\begin{array}{c}
\sigma -\lambda^2 q\\
x
\end{array}
\right),
\end{equation}
with inverse
\begin{equation}\label{eq:invartr_RT}
\left(
\begin{array}{c}
\sigma \\
x
\end{array}
\right) = \left(
\begin{array}{c c c}
\left({1 \over 2}-\alpha\right)\lambda^2 & &\left({1 \over 2}+\alpha\right)\lambda^2 \\
1 & &-1
\end{array}
\right) \left(
\begin{array}{c}
s\\
u
\end{array}
\right) + \left(
\begin{array}{c}
\lambda^2 q\\
0
\end{array}
\right),
\end{equation}
for some real $\alpha$ and $q$. Some straightforward algebra shows
that the integration domain $s=0\ldots \lambda^{-2}\tau$,
$u=0\ldots s$ in (\ref{eq:exact}), becomes the domain
$\mathcal{D}(\lambda, \tau, \alpha, q)$ in the $(\sigma, x)$ plane
given by the triangle of vertices
\begin{equation}\label{def:domain}
\mathcal{D}(\lambda, \tau, \alpha, q)=\triangle\left\{(\lambda^2
q,0), (\tau+\lambda^2 q,0),\left(\left({1\over
2}-\alpha\right)\tau+\lambda^2 q,\lambda^{-2}\tau\right)\right\}
\end{equation}
(see figure (\ref{domain1})).
Accordingly, equation (\ref{eq:exact}), written for
$W^{\lambda,i}_\tau=X_{-\lambda^{-2}\tau}
W^\lambda_{\lambda^{-2}\tau}$, becomes
\begin{eqnarray}\label{eq:taj_int}
W^{\lambda,i}_\tau &=& 1+ \iint\limits_{\mathcal{D}(\lambda, \tau,
\alpha, q)} \!\! d\sigma dx\;
X^\lambda_{-\lambda^{-2}\sigma+q-\left(\alpha+{1\over 2}\right)x}
A_{01}U^\lambda_{x} A_{10} X^\lambda_{\lambda^{-2}\sigma-q+\left(\alpha-{1\over 2}\right)x} \nonumber\\
&&\times W^{\lambda,i}_{\sigma-\lambda^2 (q + (\alpha-1/2)x)} \;.
\end{eqnarray}
We now consider the following facts, which will be made precise
and proven in Theorem \ref{th:K_(alpha_q_T)}:
\begin{itemize}
\item $\mathcal{D}(\lambda, \tau, \alpha, q)\rightarrow
[0,\tau]\times [0,\infty)$ as $\lambda\rightarrow 0$, for any real
$\alpha$ and $q$. This justifies the approximation
\begin{equation}
\iint\limits_{\mathcal{D}(\lambda, \tau, \alpha, q)}\!\!\!\!d\sigma\,dx \approx \int_0^\tau d\sigma \int_0^\infty dx \; e^{-(x/2)^2/T(\lambda)^2} \qquad \lambda\approx 0,
\end{equation}
for some real function $T$ of the coupling constant, provided
$\lim_{\lambda\rightarrow 0} T(\lambda)= +\infty$. Indeed, exponential factor would
not change the asymptotic behavior of the kernel, and would naturally conserve the original character of a bounded integration domain for every nonzero $\lambda$;

\item In the weak coupling limit $\lambda \rightarrow 0$ one could approximate
\begin{equation}\label{eq:approx_central}
W^{\lambda,i}_{\sigma-\lambda^2 (q +
(\alpha-1/2)x)} \approx W^{\lambda,i}_\sigma
\end{equation}
in the integral kernel of (\ref{eq:taj_int}).
\end{itemize}
Once all this is proven to be legitimate, equation
(\ref{eq:taj_int}) becomes
\begin{equation}
W^{\lambda,i}_\tau \approx 1+ \int_0^\tau d\sigma\;
X_{-\lambda^{-2}\sigma}^\lambda K_{(\alpha,q,T(\lambda))}
X_{\lambda^{-2}\sigma}^\lambda W^{\lambda,i}_\sigma
\end{equation}
or, which is the same,
\begin{equation}\label{eq:taj_int_appr}
W^{\lambda}_t \approx \exp\{(Z_0+\lambda A_{00}+\lambda^2 K_{(\alpha,q,T(\lambda))})t\},
\end{equation}
where we give the following
\begin{definition}\label{def:K_(alpha_q_T)}
Let $\alpha,q\in\mathbb{R}$ be two given real numbers, and let
$T\in\mathcal{C}([-1,1],\overline{\mathbb{R}})$ be a real valued positive
continuous function on the interval $[-1,1]$, where $\overline{\mathbb{R}}=\mathbb{R}\cup \{\infty\}$.
For $\lambda\in I$ define the linear operator
$K_{(\alpha,q,T(\lambda))}$ on $\mathcal{B}_0$ as
\begin{equation}
K_{(\alpha,q,T(\lambda))} = \int_0^\infty dx\; e^{-\left({x\over
2}\right)^2 / T(\lambda)^2}\; U_{-\left(\alpha+{1\over
2}\right)x+q} A_{01} U_x A_{10} U_{\left(\alpha-{1\over
2}\right)x-q}\;.
\end{equation}
Denote also with
\begin{equation}\label{def:semigr_(alpha_q)}
\overline{W}^\lambda_t=\exp\{(Z_0+\lambda A_{00}+\lambda^2
K_{(\alpha,q,T(\lambda))})t\}
\end{equation}
the associated semigroup on $\mathcal{B}_0$.
\end{definition}
Then, under fairly general hypotheses, we shall prove that
$\overline{W}^\lambda_t$ is compatible with the exact
$W^\lambda_t$ in the weak coupling limit $\lambda\rightarrow 0$,
up to $\lambda^{-2}$-rescaled positive times $\tau$:
\begin{equation}
\lim_{\lambda\rightarrow 0} \left\{ \sup_{0\leq t\leq
\lambda^{-2}\overline{\tau}}
\|W^\lambda_t-\overline{W}^\lambda_t\|\right\}=0.
\end{equation}
We stress that, as in \cite{davies2}, we \emph{do not} assume that
$\mathcal{B}_0$ is finite-dimensional.

The following lemma is not new, and for example it is contained in
Theorem 1.2 of~\cite{davies2}, but we report it here as we shall
make use of it repeatedly all throughout.
\begin{lemma}\label{lemma:volterra}
Let $b\in\mathcal{B}_0$ be given, together with some real
$\overline{\tau}>0$. Suppose $W^\lambda_t$ and
$\overline{W}^\lambda_t$ are operators on $\mathcal{B}_0$ such
that
$f_\lambda(\tau)=X^\lambda_{-\lambda^{-2}\tau}W^\lambda_{\lambda^{-2}\tau}b$
satisfies
\begin{equation}\label{eq:vns}
f_\lambda=\sum_{n\geq 0} \mathcal{H}_\lambda^n b
\end{equation}
for a Volterra operator $\mathcal{H}_\lambda$ on the Banach space
$\mathcal{V}=\mathcal{C}^0([0,\overline{\tau}],\mathcal{B}_0)$ of
continuous $\mathcal{B}_0$-valued functions on the interval
$[0,\overline{\tau}]$ (assume the same holds also for
$\overline{W}^\lambda_t$, with associated $\overline{f}_\lambda$
and $\overline{\mathcal{H}}_\lambda$). Suppose there exists a real
positive $c$ such that $\|\mathcal{H}_\lambda\|_{_\mathcal{V}}\leq
c\overline{\tau}$ and
$\|\overline{\mathcal{H}}_\lambda\|_{_\mathcal{V}}\leq
c\overline{\tau}$ uniformly on $|\lambda|\leq1$. Put
\begin{equation}
\begin{array}{ll}
i)   & \lim_{\lambda\rightarrow 0} \|\mathcal{H}_\lambda-\overline{\mathcal{H}}_\lambda\|_{_\mathcal{V}}=0\;; \\
ii)  & \lim_{\lambda\rightarrow 0} \|f_\lambda-\overline{f}_\lambda\|_{\infty}=0\;; \\
iii) & \lim_{\lambda\rightarrow 0}\; \sup_{0\leq
t\leq\lambda^{-2}\overline{\tau}}\; \|W^\lambda_t
b-\overline{W}^\lambda_t b\|_{_{\mathcal{B}_0}}=0 \;.
\end{array}
\end{equation}

Then $i)\Rightarrow ii) \Rightarrow iii)$.
\end{lemma}
{\bf Proof}. Of course $ii)\Rightarrow iii)$ as $X^\lambda_t$ is group of
isometries and one has
\begin{equation}
\sup_{0\leq t\leq \lambda^{-2}\overline{\tau}} \|W^\lambda_t b
-\overline{W}^\lambda_t b\|_{_{\mathcal{B}_0}} = \sup_{0\leq
\tau\leq
\overline{\tau}}\;\|f_\lambda(\tau)-\overline{f}_\lambda(\tau)\|_{_{\mathcal{B}_0}}=\|f_\lambda-\overline{f}_\lambda\|_\infty
\;.
\end{equation}
Then, subtracting the von Newmann expansions for $f_\lambda$ and
$\overline{f}_\lambda$ one obtains
\begin{eqnarray}
&&\|f_\lambda-\overline{f}_\lambda\|_\infty \leq \sum_{n=1}^\infty \|\mathcal{H}_\lambda^n b-\overline{\mathcal{H}}_\lambda^n b\|_{\infty} \nonumber\\
&=& \sum_{n=1}^\infty \| \mathcal{H}_\lambda^n b - \mathcal{H}_\lambda^{n-1} \overline{\mathcal{H}}_\lambda b +\mathcal{H}_\lambda^{n-1} \overline{\mathcal{H}}_\lambda b-\cdots + \mathcal{H}_\lambda \overline{\mathcal{H}}_\lambda^{n-1} b -\overline{\mathcal{H}}_\lambda^n b \|_\infty \nonumber\\
&=& \sum_{n=1}^\infty \| \mathcal{H}_\lambda^{n-1} (\mathcal{H}_\lambda-\overline{\mathcal{H}}_\lambda)b +\mathcal{H}_\lambda^{n-2} (\mathcal{H}_\lambda-\overline{\mathcal{H}}_\lambda) \overline{\mathcal{H}}_\lambda b+\cdots +(\mathcal{H}_\lambda-\overline{\mathcal{H}}_\lambda) \overline{\mathcal{H}}_\lambda^{n-1} b \|_\infty \nonumber\\
&\leq& \| \mathcal{H}_\lambda-\overline{\mathcal{H}}_\lambda
\|_{_{\mathcal{V}}} \|b\|_{_{\mathcal{B}_0}} \sum_{n=1}^\infty
{(\overline{\tau}\:c)^{n-1} \over (n-1)!}
\end{eqnarray}
and the last series is (obviously) convergent and independent of
$\lambda$. Note that we have used (and will use throughout) the
important property that if $\mathcal{H}$ is Volterra and
$\|\mathcal{H}\|\leq C$, then $\|\mathcal{H}\|^n\leq C^n/n!$.

This shows that $i)\Rightarrow ii)$ and thus finishes the proof.
$\quad\Box$ \\

We still need a technical but important result, that will allow us
to perform approximation (\ref{eq:approx_central}): its
interpretation will become clear in the context of Theorem
\ref{th:K_(alpha_q_T)}, but it deserves to be reported in the more
autonomous environment of a Lemma, as it will find application
also in our second main result, Theorem \ref{th:K_T}.
\begin{lemma}\label{lemma:mainest}
Let $\mathcal{B}_0$ be a Banach space, $\overline{\tau}>0$, and
let $\mathcal{V}=\mathcal{C}^0([0,\overline{\tau}],\mathcal{B}_0)$
be the Banach space of continuous functions from
$[0,\overline{\tau}]$ into $\mathcal{B}_0$. For some real $q$ and
$\alpha$, let $\mathcal{D}(\lambda, \tau, \alpha, q)$ be the
triangle in the $(\sigma,x)$-plane of vertices
\begin{equation}
\mathcal{D}(\lambda, \tau, \alpha, q)=\triangle\left\{(\lambda^2
q,0), (\tau+\lambda^2 q,0),\left(\left({1\over
2}-\alpha\right)\tau+\lambda^2 q,\lambda^{-2}\tau\right)\right\}
\end{equation}
and define the truncated domain (see figure
(\ref{domain1}))
\begin{equation}\label{def:domain_truncated}
\overline{\mathcal{D}}(\lambda, \tau, \alpha,
q)=\mathcal{D}(\lambda, \tau, \alpha, q)\cap [0,\tau]\times
[0,\infty).
\end{equation}
Let $T\in\mathcal{C}([-1,1],\overline{\mathbb{R}})$ be a real valued positive
continuous function on the interval $[-1,1]$, and assume
that $T(\lambda)\sim|\lambda|^\xi$ for $\lambda\rightarrow 0$ with
$\xi<2$. Let $\mathcal{H}_{(\lambda \alpha q)}$ be a Volterra
integral operator on $\mathcal{V}$ and assume it can be put in the
form
\begin{equation}\label{eq:var_change}
(\mathcal{H}_{(\lambda \alpha q)}
g)(\tau)=\int\!\!\!\!\!\!\!\!\!\!\!\!\int\limits_{\overline{\mathcal{D}}(\lambda,
\tau, \alpha, q)} \!\!\!\!\!\! d\sigma dx \; e^{-{(x/ 2)^2\over
T(\lambda)^2}}\; K_{\alpha q}(\lambda,\sigma,x)\:
g\left(\sigma-\lambda^2 \left(q + \left(\alpha-{1/
2}\right)x\right)\right),
\end{equation}
for a suitable kernel $K_{\alpha q}(\lambda,\sigma,x)$, and define
the Volterra integral operator $\overline{\mathcal{H}}_{(\lambda
\alpha q)}$ by
\begin{equation}
(\overline{\mathcal{H}}_{(\lambda \alpha q)}
g)(\tau)=\int\!\!\!\!\!\!\!\!\!\!\!\!\int\limits_{\overline{\mathcal{D}}(\lambda,
\tau, \alpha, q)} \!\!\!\!\!\! d\sigma dx \; e^{-{(x/ 2)^2\over
T(\lambda)^2}}\; K_{\alpha q}(\lambda,\sigma,x)\: g(\sigma).
\end{equation}
Suppose that $\| K_{\alpha q}(\lambda,\sigma,x) \|=k(x)$
independently on $\sigma,\lambda,\alpha,q$, and that
\begin{equation}\label{hp:bound_2}
\int_0^\infty dx\;k(x)=c
\end{equation}
for some finite $0<c<\infty$, and assume $\|\mathcal{H}_{(\lambda
\alpha q)}\|<\overline{\tau}c$.

Let $b\in\mathcal{B}_0$ and define the von Neumann series
$f_{(\lambda \alpha q)}$ and $\overline{f}_{(\lambda \alpha q)}$
through equation (\ref{eq:vns}) in Lemma (\ref{lemma:volterra}).

Then
\begin{equation}\label{eq:estriv}
\|\overline{\mathcal{H}}_{(\lambda \alpha
q)}\|\leq\overline{\tau}c
\end{equation}
and
\begin{equation}\label{eq:estnontriv}
\lim_{\lambda\rightarrow 0} \|f_{(\lambda \alpha q)} -
\overline{f}_{(\lambda \alpha q)}\|_\infty =0 \;.
\end{equation}
\end{lemma}

{\bf Proof}.
Estimation (\ref{eq:estriv}) is a trivial consequence of the
definition of $\overline{\mathcal{H}}_{(\lambda \alpha q)}$, as
\begin{equation}
\|\overline{\mathcal{H}}_{(\lambda \alpha q)}g(\tau)\|\leq \tau
\int_0^\infty k(x) \|g\|_\infty.
\end{equation}
To show the validity of (\ref{eq:estnontriv}) we start proceeding
as in Lemma~\ref{lemma:volterra} to obtain
\begin{equation}\label{lemma:mainest:eq:1}
\|f_{(\lambda \alpha q)}-\overline{f}_{(\lambda \alpha q)}\|_\infty \nonumber \\
\leq \sum_{n=1}^\infty\; \sum_{l=1}^{n-1}\;
{(\overline{\tau}\:c)^{n-l-1} \over (n-l)!} \;
\|(\mathcal{H}_{(\lambda \alpha
q)}-\overline{\mathcal{H}}_{(\lambda \alpha
q)})\:\overline{\mathcal{H}}_{(\lambda \alpha q)}^l\: b\|_\infty
\;.
\end{equation}
Note that the case $l=0$ has been dropped, since
$\mathcal{H}_{(\lambda \alpha
q)}b=\overline{\mathcal{H}}_{(\lambda \alpha q)} b$ trivially, as
$b$ is constant, as one can see from (\ref{eq:var_change}). Now if
we could show that for every $\epsilon>0$ there exists some
$\overline{\lambda}>0$ such that $|\lambda|<\overline{\lambda}$
implies that for every $l\geq 1$
\begin{equation}\label{eq:sufficient_lemma}
\|(\mathcal{H}_{(\lambda \alpha
q)}-\overline{\mathcal{H}}_{(\lambda \alpha
q)})\:\overline{\mathcal{H}}_{(\lambda \alpha q)}^l\: b\|_\infty
\leq {(\overline{\tau}\:c)^{l-1} \over (l-1)!} \:
\overline{\tau}\; \|b\|_{\mathcal{B}_0}  \: \epsilon \;,
\end{equation}
we would be done, as, following estimation
(\ref{lemma:mainest:eq:1}), we would have
\begin{equation}
\|f_\lambda-\overline{f}_\lambda\|_\infty \leq \sum_{n=1}^\infty\;
\sum_{l=1}^{n-1}\; {(\overline{\tau}\:c)^{n-2} \over (n-l)!(l-1)!}
\: \overline{\tau}\; \|b\|_{\mathcal{B}_0}  \: \epsilon \;,
\end{equation}
and the series would obviously converge (because $c>0$ by
hypothesis).
\begin{figure}[htb]
\includegraphics[width=9cm,angle=0]{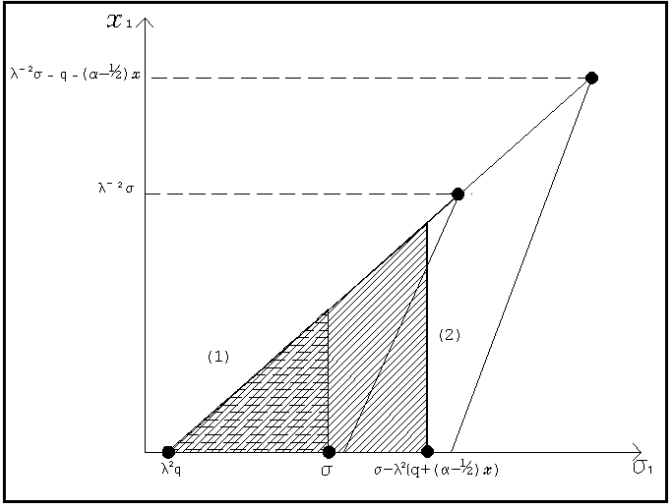}
\caption{Integration domains $\overline{\mathcal{D}}(\lambda, \sigma, \alpha, q)$
(rectangular triangle $(1)$), $\overline{\mathcal{D}}(\lambda, \sigma-\lambda^2
\left(q + \left(\alpha-{1\over 2}\right)x\right), \alpha, q)$
(rectangular triangle $(2)$) and $S(\lambda,\sigma,x)$ (region
$(1)\cup(2)$). We have put $\alpha<-1/2$, $q>0$ for clarity.
The two truncated triangles $\overline{\mathcal{D}}$ are similar, as are also the untruncated, underlying $\mathcal{D}$. Note the scaling behavior with $\lambda^2$. \label{fig:domain2}}
\end{figure}
To show that property (\ref{eq:sufficient_lemma}) holds, we take
$l>0$ and evaluate
\begin{eqnarray}
&& \left[(\mathcal{H}_{(\lambda \alpha q)}-\overline{\mathcal{H}}_{(\lambda \alpha q)})\:\overline{\mathcal{H}}_{(\lambda \alpha q)}^l\: b\right](\tau) \nonumber \\
&=& \iint\limits_{\overline{\mathcal{D}}(\lambda, \tau, \alpha, q)} \!\!\!\!\!\! d\sigma dx \; e^{-{(x/ 2)^2\over T(\lambda)^2}}\; K_{\alpha q}(\lambda,\sigma,x) \nonumber\\
&&\times \left\{ (\overline{\mathcal{H}}_{(\lambda \alpha q)}^l\: b)\left(\sigma-\lambda^2 (q +( \alpha-1/2)x)\right)-(\overline{\mathcal{H}}_{(\lambda \alpha q)}^l\: b)(\sigma) \right\} \nonumber \\
&=&  \iint\limits_{\overline{\mathcal{D}}(\lambda, \tau, \alpha, q)}  d\sigma dx \; e^{-{(x/ 2)^2\over T(\lambda)^2}}\; K_{\alpha q}(\lambda,\sigma,x) \nonumber\\
&&\times\iint\limits_{S(\lambda,\alpha,q,\sigma,x)} d\sigma_1 dx_1
\, \Delta^\lambda(\sigma_1,x_1)\,e^{-{(x/ 2)^2\over
T(\lambda)^2}}\; K_{\alpha q}(\lambda,\sigma_1,x_1)\,
(\overline{\mathcal{H}}_{(\lambda \alpha q)}^{l-1} b)(\sigma_1)
\end{eqnarray}
where we have put $\Delta^\lambda=\chi^\lambda_1-\chi^\lambda_2$,
$\chi^\lambda_1$ being the characteristic function of
$\overline{\mathcal{D}}(\lambda, \sigma-\lambda^2 \left(q +
\left(\alpha-{1\over 2}\right)x\right), \alpha, q)$,
$\chi^\lambda_2$ the characteristic function of
$\overline{\mathcal{D}}(\lambda, \sigma, \alpha, q)$, and we have
defined (see Figure \ref{fig:domain2})
\begin{equation}
S(\lambda,\alpha,q,\sigma,x)=\overline{\mathcal{D}}\left(\lambda,
\sigma-\lambda^2 \left(q + \left(\alpha-{1\over 2}\right)x\right),
\alpha, q\right)\cup \overline{\mathcal{D}}(\lambda, \sigma,
\alpha, q)\;.
\end{equation}
Passing to the norms, we estimate the left hand side of
(\ref{eq:sufficient_lemma}) as
\begin{eqnarray}\label{est:mainlemma}
&&\|(\mathcal{H}_\lambda-\overline{\mathcal{H}}_{(\lambda \alpha q)})\:\overline{\mathcal{H}}_{(\lambda \alpha q)}^l\: b\|\nonumber\\
&\leq& \sup_{0\leq \tau \leq \overline{\tau}} \iint\limits_{\overline{\mathcal{D}}(\lambda, \tau, \alpha, q)} \!\!\!\!\!\! d\sigma dx \, e^{-{(x/ 2)^2\over T(\lambda)^2}}\,k(x)\iint\limits_{S(\lambda,\alpha,q,\sigma,x)} \!\!\!d\sigma_1 dx_1 \, |\Delta^\lambda(\sigma_1,x_1)|\,k(x_1)\,\left\|\overline{\mathcal{H}}_{(\lambda \alpha q)}^{l-1} b\right\|_\infty \nonumber\\
&\leq&{(\overline{\tau}\:c)^{l-1} \over (l-1)!} \:\|b\|_{\mathcal{B}_0} \sup_{0\leq \tau \leq \overline{\tau}}  \iint\limits_{\overline{\mathcal{D}}(\lambda, \tau, \alpha, q)} \!\!\!\!\!\! d\sigma dx \,e^{-{(x/ 2)^2\over T(\lambda)^2}}\, k(x) \; \Xi(\lambda,\sigma,x) \, \nonumber \\
&\leq&{(\overline{\tau}\:c)^{l-1} \over (l-1)!} \:\|b\|_{\mathcal{B}_0} \overline{\tau} \sup_{0\leq \sigma \leq \overline{\tau}} \int_0^{\infty} dx \; e^{-{(x/ 2)^2\over T(\lambda)^2}}k(x) \; \Xi_{\alpha q}(\lambda,\sigma,x) \,\nonumber\\
\end{eqnarray}
where we named
\begin{equation}\label{def:Xi}
\Xi_{\alpha q}(\lambda,\sigma,x)
=\iint\limits_{S(\lambda,\alpha,q,\sigma,x)} \!\!\!d\sigma_1 dx_1
\, |\Delta^\lambda(\sigma_1,x_1)|\,k(x_1)
\end{equation}
At this point let us consider the sector defined by $q>0$ and
$\alpha<-1/2$. In this case, it is but a straightforward algebra
to show that the two triangles ${\mathcal{D}}\left(\lambda,
\sigma-\lambda^2 \left(q + \left(\alpha-{1\over 2}\right)x\right),
\alpha, q\right)$ and ${\mathcal{D}}(\lambda, \sigma, \alpha, q)$
(the same applies to $\overline{\mathcal{D}}$) are similar, and
that their left edges lie on the same line of equation
\begin{equation}
\overline{x}_l(\lambda,\sigma_1)={\lambda^{-2}\over {1\over
2}-\alpha}(\sigma_1-\lambda^2 q)
\end{equation}
in the $(\sigma_1,x_1)$-plane (see figure (\ref{fig:domain2})). In
particular,
\begin{equation}
\overline{\mathcal{D}}\left(\lambda, \sigma-\lambda^2 \left(q +
\left(\alpha-{1\over 2}\right)x\right), \alpha, q\right)\;
\begin{array}{c}
\supseteq \\
\subset
\end{array}
\;\overline{\mathcal{D}}(\lambda, \sigma, \alpha, q), \quad x\gtreqless{q\over {1\over 2}-\alpha} \nonumber\\
\end{equation}
For this reason we easily compute
\begin{eqnarray}
&&\Xi_{\alpha q}(\lambda,\sigma,x)={\rm{sign}}\left(\left({1\over 2}-\alpha\right)x-q\right) \int_\sigma^{\sigma-\lambda^2\left(q+\left(\alpha-{1\over 2}\right)x\right)}\!\!\!\!\!\!\!\!\!\!\!\! d\sigma_1 \int_0^{\overline{x}_l(\lambda,\sigma_1)}\!\!\!\!\!\! dx_1\; k(x_1) \nonumber\\
&&\leq \lambda^2 \left|q+\left(\alpha-{1\over 2}\right)x\right| \int_0^\infty dx_1\; k(x_1) \nonumber\\
&&\leq c \lambda^2 \left(|q|+\left|\alpha-{1\over 2}\right|x
\right)
\end{eqnarray}
According to the last line in estimation (\ref{est:mainlemma}), we
study
\begin{eqnarray}\label{eq:ninfa}
&&\sup_{0\leq \sigma \leq \overline{\tau}} \int_0^{\infty} dx \; e^{-{(x/ 2)^2\over T(\lambda)^2}}k(x) \; \Xi_{\alpha q} (\lambda,\sigma,x)\nonumber\\
&& \leq c^2\lambda^2 q + c\left|\alpha-{1\over 2}\right|
\int_0^{\infty} dx \; k(x) e^{-{(x/ 2)^2\over
T(\lambda)^2}}\lambda^2 x
\end{eqnarray}
The first term goes to zero with velocity $\sim \lambda^2 q$,
whereas the dominated convergence theorem applies to the second
term, showing convergence to zero uniformly on $q$, with velocity
$\sim \lambda^{2-\xi}$ (note that we have supposed $\xi<2$
strictly). This, when put in (\ref{est:mainlemma}), shows the
validity of (\ref{eq:sufficient_lemma}).

A very similar analysis can be done for the remaining sectors
$q\lesseqgtr 0$, $|\alpha|<1/2$ and $\alpha>1/2$. In each case,
the net result is that the estimation (\ref{eq:ninfa}) is always
of the form
\begin{equation}\label{eq:ninfa_general}
\sup_{0\leq \sigma \leq \overline{\tau}} \int_0^{\infty} dx \;
e^{-{(x/ 2)^2\over T(\lambda)^2}}k(x) \; \Xi_{\alpha
q}(\lambda,\sigma,x) \;\leq\; C_1\lambda^2 |q| + C_2
\lambda^{2-\xi}
\end{equation}
for suitable real positive constants $C_1$ and $C_2$. Again, this
shows the validity of (\ref{eq:sufficient_lemma}), and thus
concludes the proof.
$\quad\Box$ \\

We are now in position to state our first main result:
\begin{theorem}\label{th:K_(alpha_q_T)}
Suppose that $X^\lambda_t$ is a one-parameter group of isometries.

Let $T\in\mathcal{C}([-1,1],\overline{\mathbb{R}})$ be a real valued positive
continuous function on the interval $[-1,1]$, and assume
\begin{equation}
T(\lambda)\sim |\lambda|^{-\xi} \widetilde{T} ,\quad
\lambda\rightarrow 0
\end{equation}
for some real positive $\widetilde{T}>0$ and $0<\xi<2$ (strictly).

Suppose that there exists some $0<c<\infty$ such that for every
$\overline{\tau}>0$
\begin{equation}\label{eq:no_oscillation_lambda}
\int_0^{\lambda^{-2}\overline{\tau}} \|A_{01}U^\lambda_x A_{10}\|
\, dx \leq c
\end{equation}
is bounded uniformly on $|\lambda|\leq 1$.

Suppose also that for every $0<\overline{\tau}<\infty$
\begin{equation}\label{hp:lambda_convergence}
\lim_{\lambda\rightarrow 0} \int_0^{\lambda^{-2}\overline{\tau}}
\|A_{01}(U^\lambda_x - U_x) A_{10}\| \, dx = 0 \;.
\end{equation}

Then for every $\overline{\tau}>0$
\begin{equation}
\lim_{\lambda\rightarrow 0} \left\{ \sup_{0\leq t\leq
\lambda^{-2}\overline{\tau}}
\|W^\lambda_t-\overline{W}^\lambda_t\|\right\}=0.
\end{equation}
\end{theorem}

{\bf Proof}.
Let $\mathcal{V}$ be the Banach space of norm continuous
$\mathcal{B}_0$-valued functions on $[0,\overline{\tau}]$, and let
$b\in\mathcal{B}_0$. Define the "interaction picture" time
rescaled solution of (\ref{eq:exact})
\begin{equation}
f_\lambda(\tau)=X^\lambda_{-\lambda^{-2}\tau}W^\lambda_{\lambda^{-2}\tau}b.
\end{equation}
Then $f_\lambda$ is a solution to the integral equation
\begin{equation}
f_\lambda=b+\mathcal{H}_\lambda f_\lambda,
\end{equation}
where the integral operator $\mathcal{H}_\lambda$ is defined
(recall that $X^\lambda_t$ is a group of isometries) by
\begin{equation}\label{eq:intopexact}
(\mathcal{H}_\lambda g)(\tau)=\lambda^2
\int_0^{\lambda^{-2}\tau} \!\!\!\! ds \int_0^s  du \;
X^\lambda_{s}A_{01}U^\lambda_{s-u} A_{10} X^\lambda_{u}\;
g(\lambda^2 u) \;.
\end{equation}
In \cite{davies2} Davies shows that $\mathcal{H}_\lambda$ is a
Volterra operator. Indeed, by changing coordinates according to
(\ref{eq:vartr_davies}), Eq. (\ref{eq:intopexact}) can be given an explicit Volterra form,
namely as
\begin{equation}\label{eq:volterra}
(\mathcal{H}_\lambda g)(\tau)=\int_0^{\tau}  d\sigma
X^\lambda_{-\lambda^{-2}\sigma} K(\lambda,\tau-\sigma)
X^\lambda_{\lambda^{-2}\sigma}\; g(\sigma) \;,
\end{equation}
where we defined the "slowly varying" kernel
\begin{equation}
K(\lambda,\tau)= \int_0^{\lambda^{-2}\tau} dx \; X^\lambda_{-x}
A_{01} U^\lambda_{x} A_{10}.
\end{equation}
Because of this reason, and since $K(\lambda,\tau)$ is manifestly
bounded by $c$, uniformly on $\lambda$ (thanks our boundedness
hypothesis), it follows that
\begin{equation}\label{eq:volt_power}
\|\mathcal{H}_\lambda^n\|\leq c^n\overline{\tau}^n/n!,
\end{equation}
and also that the associated von Newmann series expansion
\begin{equation}\label{eq:von_newmann}
f_\lambda=b+\mathcal{H}_\lambda b+\mathcal{H}_\lambda^2 b+\cdots
\end{equation}
converges.

We can proceed in similar fashion also for the semigroup
(\ref{def:semigr_(alpha_q)}): iteration gives indeed
\begin{equation}
\overline{W}^\lambda_{\lambda^{-2}\tau}=X^\lambda_{-\lambda^{-2}\tau}+
\int_0^\tau d\sigma\; X^\lambda_{-\lambda^{-2}(\tau-\sigma)}
K_{(\alpha,q,T(\lambda))} W^\lambda_{\lambda^{-2}\sigma} \;.
\end{equation}
Accordingly, we define
\begin{equation}
\overline{f}_\lambda(\tau)=X^\lambda_{-\lambda^{-2}\tau}\overline{W}^\lambda_{\lambda^{-2}\tau}b
\end{equation}
so that it follows that $\overline{f}_\lambda$ is a solution to
the integral equation
\begin{equation}
\overline{f}_\lambda=b+\overline{\mathcal{H}}_{(\lambda\alpha q)}
\overline{f}_\lambda,
\end{equation}
where we have defined
\begin{equation}\label{eq:intop}
(\overline{\mathcal{H}}_{(\lambda\alpha q)} g)(\tau)=\int_0^\tau
d\sigma\; X^\lambda_{-\lambda^{-2}\sigma}
K_{(\alpha,q,T(\lambda))} X^\lambda_{\lambda^{-2}\sigma}
g(\sigma).
\end{equation}
Now again, $\overline{\mathcal{H}}_{(\lambda\alpha q)}$ is a
Volterra operator, and since $\|K_{\alpha,q}\| \leq c$, equations
(\ref{eq:volt_power}) and (\ref{eq:von_newmann}) follow
analogously for $\overline{\mathcal{H}}_{(\lambda\alpha q)}$.

Clearly (see Lemma \ref{lemma:volterra}), we must show that for
any chosen $b\in\mathcal{B}_0$,
\begin{equation}
\sup_{0\leq t\leq \lambda^{-2}\overline{\tau}} \|W^\lambda_t b
-\overline{W}^\lambda_t b\|
=\|f_\lambda-\overline{f}_\lambda\|_\infty \rightarrow 0, \quad
\lambda\rightarrow 0.
\end{equation}
We shall do that by defining suitable Volterra operators, denoted
with $\mathcal{H}^{(j)}_{(\lambda\alpha q)}$, $j=0\ldots N$, such
that $\mathcal{H}^{(0)}_{(\lambda\alpha q)}=\mathcal{H}_\lambda$,
$\mathcal{H}^{(N)}_{(\lambda\alpha
q)}=\overline{\mathcal{H}}_{(\lambda\alpha q)}$, and either we can
show, according to Lemma \ref{lemma:volterra}, that
\begin{equation}
\| \mathcal{H}^{(j)}_{(\lambda\alpha q)} -
\mathcal{H}^{(j-1)}_{(\lambda\alpha q)} \|_{_\mathcal{V}}
\rightarrow 0 \;,\quad \lambda\rightarrow 0
\end{equation}
or more directly that
\begin{equation}
\|f^{(j)}_{(\lambda\alpha q)}-f^{(j-1)}_{(\lambda\alpha
q)}\|_\infty \rightarrow 0\;,\quad \lambda\rightarrow 0
\end{equation}
where
\begin{equation}
f^{(j)}_{(\lambda\alpha q)}=\sum_{n=0}^\infty
\left(\mathcal{H}^{(j)}_{(\lambda\alpha q)}\right)^n b
\end{equation}
is the associated von Neumann series. Then, our conclusion (see
Lemma \ref{lemma:volterra}) would follow from\footnote{Note that
we have attached the subscript "$(\lambda\alpha q)$" all
throughout: although some Volterra operator may not actually
depend on $\alpha$, nor $q$, this unifying notation will become
useful in the sequel.}
\begin{equation}\label{eq:proof_1_conclusion}
\|f_\lambda-\overline{f}_\lambda\| \leq \sum_{j=1}^N \|
f^{(j)}_{(\lambda\alpha q)} - f^{(j-1)}_{(\lambda\alpha q)} \|
\rightarrow 0 \quad\lambda\rightarrow 0.
\end{equation}

To follow our purpose, instead of (\ref{eq:vartr_davies}), we
perform the coordinate transformations (\ref{eq:vartr_RT})
As we noted before, the integration domain $s=0\ldots
\lambda^{-2}\tau$, $u=0\ldots s$, becomes the domain
$\mathcal{D}(\lambda, \tau, \alpha, q)$ in the $(\sigma, x)$-plane
defined in (\ref{def:domain}), and depicted in figure
(\ref{domain1}). Accordingly, the integral kernel in (\ref{eq:intopexact}) is now
written as
\begin{eqnarray}\label{eq:intopexactsigmax}
(\mathcal{H}_\lambda g)(\tau)&=&\iint\limits_{\mathcal{D}(\lambda, \tau, \alpha, q)} \!\! d\sigma dx\;  X^\lambda_{-\lambda^{-2}\sigma+q-\left(\alpha+{1\over 2}\right)x} A_{01}U^\lambda_{x} A_{10} X^\lambda_{\lambda^{-2}\sigma-q+\left(\alpha-{1\over 2}\right)x} \nonumber\\
&&\times g(\sigma-\lambda^2 (q + (\alpha-{1/2})x)) \;.
\end{eqnarray}

We shall work throughout with the choice $\alpha<-1/2$ and $q>0$,
without loss of generality (the structure of the proof is
basically the same for all the remaining sectors, which we leave
to the reader, and actually the case $|\alpha|<1/2$ presents less
difficulties).

The first thing we shall be concerned with, is to find a way to
substitute the free "polarization evolution" $P_1 U_x$ in place of
the interacting $P_1 U^\lambda_x$ in the middle of the kernel in
(\ref{eq:intopexactsigmax}). This step is accomplished by defining
a related integral operator $\mathcal{H}^{(1)}_\lambda$ by
\begin{eqnarray}\label{def:h1}
(\mathcal{H}^{(1)}_{(\lambda\alpha q)} g)(\tau)&=&\iint\limits_{\mathcal{D}_{(1)}(\lambda, \tau, \alpha, q)} \!\!\! d\sigma dx\;  X^\lambda_{-\lambda^{-2}\sigma+q-\left({\alpha+1\over 2}\right)x} A_{01}U_x A_{10} X^\lambda_{\lambda^{-2}\sigma-q+\left(\alpha-{1\over 2}\right)x} \nonumber\\
&\times& g(\sigma-\lambda^2 (q + (\alpha-{1/2})x)) \;.
\end{eqnarray}
Here we have denoted $\mathcal{D}_{(1)}=\mathcal{D}$ for sake of
notation. Note that $\mathcal{H}^{(1)}_{(\lambda\alpha q)}$ does
not depend on $\alpha$, nor on $q$, although both the latter
parameters appear in its definition. To compare the two integral
operators, take a bounded $g\in\mathcal{V}$ and estimate
\begin{eqnarray}
&& \left\|\:\left(\mathcal{H}_\lambda-\mathcal{H}^{(1)}_{(\lambda\alpha q)}\right)g \: \right\| \;\leq \; \iint\limits_{\mathcal{D}(\lambda, \tau, \alpha, q)} \!\!\! d\sigma dx\;  \| A_{01}(U^\lambda_x-U_{x}) A_{10} \| \; \|g\|_\infty \nonumber \\
&&\leq \max\{1,|1/2-\alpha|\}\; \overline{\tau} \|g\|_\infty \int_{0}^{\lambda^{-2}\overline{\tau}}\!\!\!\!\! dx \;  \| A_{01}(U^\lambda_x-U_{x}) A_{10} \|
\end{eqnarray}
(note for later purposes that the estimation does not depend on
$q$). Our convergence hypothesis (\ref{hp:lambda_convergence}) on
$U^\lambda_x$ allows then to conclude that this goes to zero when
$\lambda\rightarrow 0$ uniformly on every $\|g\|=1$, so that we
obtain
\begin{equation}\label{eq:estimation_1}
\lim_{\lambda\rightarrow
0}\left\|\:\mathcal{H}_\lambda-\mathcal{H}^{(1)}_{(\lambda\alpha
q)} \: \right\| =0 \;.
\end{equation}
We proceed along similar lines to smooth the kernel with
$T(\lambda)$: define
\begin{eqnarray}\label{def:h2}
(\mathcal{H}^{(2)}_{(\lambda\alpha q)} g)(\tau)&=&\!\!\!\iint\limits_{\mathcal{D}_{(2)}(\lambda, \tau, \alpha, q)} \!\!\!\!\! d\sigma dx\;  e^{-{(x/ 2)^2 \over T(\lambda)^2}}\; X^\lambda_{-\lambda^{-2}\sigma+q-\left(\alpha+{1\over 2}\right)x} A_{01}U_x A_{10} X^\lambda_{\lambda^{-2}\sigma-q+\left(\alpha-{1\over 2}\right)x} \nonumber\\
&\times& g(\sigma-\lambda^2 (q + (\alpha-{1/2})x)) \;.
\end{eqnarray}
Here we have denoted $\mathcal{D}_{(2)}=\mathcal{D}_{(1)}$. To
compare with $\mathcal{H}^{(1)}_{(\lambda\alpha q)}$, take a
bounded $g\in\mathcal{V}$ and estimate
\begin{eqnarray}
&& \left\|\:\left(\mathcal{H}^{(1)}_{(\lambda\alpha q)}-\mathcal{H}^{(2)}_{(\lambda\alpha q)}\right)g \: \right\| \leq \iint\limits_{\mathcal{D}(\lambda, \tau, \alpha, q)} \!\!\! d\sigma dx\; |1-e^{-({x\over 2})^2/T(\lambda)^2}|\; \| A_{01}U_x A_{10} \| \; \|g\|_\infty \nonumber \\
&&\leq \max\left\{1,\left|{1\over 2}-\alpha\right|\right\}\;\overline{\tau} \|g\|_\infty \int_{0}^{\infty} dx \; |1-e^{-{(x/ 2)^2 \over T(\lambda)^2}}|\; \| A_{01}U_x A_{10} \|
\end{eqnarray}
Hypothesis (\ref{eq:no_oscillation_lambda}) furnishes an
integrable upper bound to the last integrand, and so the integral
goes to zero in the limit $\lambda\rightarrow 0$ because of the
dominated convergence theorem: in fact, one has pointwise
convergence
\begin{equation}
\left|1-e^{-{(x/ 2)^2 \over T(\lambda)^2}}\right|
\rightarrow 0, \quad \lambda\rightarrow 0\;
\end{equation}
due to our hypothesis $0<\xi$. Uniform convergence on all
$\|g\|=1$ in the last line of our estimation then shows that
\begin{equation}\label{eq:estimation_2}
\lim_{\lambda\rightarrow
0}\left\|\:\mathcal{H}^{(1)}_{(\lambda\alpha
q)}-\mathcal{H}^{(2)}_{(\lambda\alpha q)} \: \right\| =0 \;.
\end{equation}
We note that $\mathcal{H}^{(2)}_{(\lambda\alpha q)}$, as
$\mathcal{H}^{(1)}_{(\lambda\alpha q)}$ and $\mathcal{H}_\lambda$,
is also a Volterra operator, as is easy to verify: it would
suffice to apply the inverse transform (\ref{eq:invartr_RT}) for
the specific choice of $\alpha$ and $q$, and subsequently apply
the transform (\ref{eq:vartr_RT}) for $q=0$ and $\alpha=1/2$, to
find $\mathcal{H}^{(2)}_{(\lambda\alpha q)}$ into its explicit
Volterra form (and also find that it does not depend on
$(\alpha,q)$).

Now define the restricted domain (see figure (\ref{domain1}))
\begin{equation}
\mathcal{D}_{(3)}(\lambda, \tau, \alpha,
q)=\mathcal{D}_{(2)}(\lambda, \tau, \alpha, q)\;\cap\;
[0,\tau]\times [0,\infty)\;
\end{equation}
and define $\mathcal{H}^{(3)}_{(\lambda \alpha q)}$ accordingly,
as the restriction of $\mathcal{H}^{(2)}_{(\lambda \alpha q)}$ to
$\mathcal{D}_{(3)}(\lambda, \tau, \alpha, q)$, that is,
\begin{eqnarray}\label{def:h3}
(\mathcal{H}^{(3)}_{(\lambda \alpha q)} g)(\tau)&=&\!\!\!\!\iint\limits_{\mathcal{D}_{(3)}(\lambda, \tau, \alpha, q)}
\!\!\!\!\!\! d\sigma dx\; e^{-{(x/ 2)^2 \over T(\lambda)^2}}\; X^\lambda_{-\lambda^{-2}\sigma+q-\left(\alpha+{1\over 2}\right)x} A_{01}U_x A_{10} X^\lambda_{\lambda^{-2}\sigma-q+\left(\alpha-{1\over 2}\right)x} \nonumber\\
&\times& g(\sigma-\lambda^2 (q + (\alpha-{1/2})x)) \;.
\end{eqnarray}
Of course this is again a Volterra operator, as the image of
$\mathcal{D}_{(3)}(\lambda, \tau, \alpha, q)$ under the
composition of (\ref{eq:invartr_RT}) and (\ref{eq:vartr_RT}), for
$q=0$ and $\alpha=1/2$, is inside the rectangular triangle, image
of $\mathcal{D}_{(2)}(\lambda, \tau, \alpha, q)$ through the same
transformations. So the composition of (\ref{eq:invartr_RT}) and
(\ref{eq:vartr_RT}), for $q=0$ and $\alpha=1/2$, will put both
$\mathcal{H}^{(2)}_{(\lambda \alpha q)}$ and
$\mathcal{H}^{(3)}_{(\lambda \alpha q)}$ in their explicit
Volterra form.

We must prove that
\begin{equation}\label{eq:estimation_3}
\lim_{\lambda\rightarrow 0} \left\|\:\mathcal{H}^{(2)}_{(\lambda
\alpha q)}-\mathcal{H}^{(3)}_{(\lambda \alpha q)} \: \right\|
\rightarrow 0 \;
\end{equation}
(note that for $|\alpha|<1/2$ this is obvious, as one has
$\mathcal{H}^{(2)}_{(\lambda \alpha
q)}=\mathcal{H}^{(3)}_{(\lambda \alpha q)}$, for the upper vertex
of the triangular domain projects on the triangle $\sigma$-axis
basis, and thus $\mathcal{D}_{(2)}(\lambda, \tau, \alpha,
q)=\mathcal{D}_{(3)}(\lambda, \tau, \alpha, q)$). To this end, we
take as usual any $g\in\mathcal{V}$ with $\|g\|=1$ and estimate
\begin{eqnarray}
&&\left\|(\left(\mathcal{H}^{(2)}_\lambda-\mathcal{H}^{(3)}_{(\lambda
\alpha q)}\right)g)(\tau) \right\|  \leq
\!\!\!\!\!\!\!\!\!\!\!\!\!\!\!\!\!\!\!\!\!\!\!\!\!\!\!\!\!\!\!\!\!\!\!\!\!\!\!\!\!\!\!\!\!\!\!\!\!
\iint\limits_{\quad\qquad\qquad\qquad\qquad\mathcal{D}(\lambda,
\tau, \alpha, q)\cap \{\tau\leq\sigma\leq
|1/2-\alpha|\tau+\lambda^2q\}}
\!\!\!\!\!\!\!\!\!\!\!\!\!\!\!\!\!\!\!\!\!\!\!\!\!\!\!\!\!\!\!\!\!\!\!\!\!\!\!\!\!\!\!\!\!\!\!\!\!\!\!
d\sigma dx\;  \| A_{01}U_{x} A_{10} \| \; \|g\|_\infty \nonumber \\
&&\leq \|g\|_\infty \left\{
\lambda^2 q \int_{0}^{\infty}\!\! dx \;  \| A_{01}U_{x} A_{10} \|+\int_{\tau+\lambda^2 q}^{\left({1\over 2}-\alpha\right)\tau+\lambda^2 q}\!\!\!\!\!\!\!\!\!\! d\sigma \int_{\overline{x}_r(\lambda,q,\sigma)}^{\infty} \!\!\!\!\!\! dx \;  \| A_{01}U_{x} A_{10} \| \right\} \nonumber\\
\end{eqnarray}
Here
\begin{equation}
\overline{x}_r(\lambda,\sigma)={q+\lambda^{-2}(\tau-\sigma) \over
\alpha+{1\over 2}}
\end{equation}
is the $x$-coordinate of the right edge of the triangle
$\mathcal{D}(\lambda, \tau, \alpha, q)$ (see figure
(\ref{domain1})). The first term in the curly brackets clearly
goes to zero uniformly on $\tau$ (with speed $\sim
\lambda^{-2}q$), because of our boundedness hypothesis
(\ref{eq:no_oscillation_lambda}). In the second term in the curly
brackets, we change coordinates according to $\sigma\rightarrow
\sigma-\lambda^2 q$, so that it becomes equal to
\begin{equation}
\int_{\tau}^{\left({1\over 2}-\alpha\right)\tau}\!\!\!\!\!\!
d\sigma \int_{\overline{x}_r(\lambda,0,\sigma)}^{+\infty}
\!\!\!\!\!\!dx \;  \| A_{01}U_{x} A_{10} \| .
\end{equation}
Now since $\overline{x}_r(\lambda,0,\sigma)>0$ a.e. in the
$\sigma$-integration domain (everywhere except $\sigma=\tau$), we
have the following convergence (pointwise with respect to
$\sigma$)
\begin{equation}
\lim_{\lambda\rightarrow 0}
\int_{\overline{x}_r(\lambda,0,\sigma)}^{+\infty} dx \;  \|
A_{01}U_{x} A_{10} \|= 0\;,
\end{equation}
thanks to hypothesis (\ref{eq:no_oscillation_lambda}) for
$\lambda=0$. Since the integration domain for $\sigma$ is bounded,
convergence to zero follows for the first integral in the curly
brackets in the estimation above, because of the dominated
convergence theorem. Uniform convergence on all
$0\leq\tau\leq\overline{\tau}$ (and real $q$) follows from the
fact that the $\sigma$-integration domain is compact.  This proves
convergence (\ref{eq:estimation_2}).

According to our roadmap, we now define the following "time
localized" Volterra operator:
\begin{equation}\label{def:h4}
(\mathcal{H}^{(4)}_{(\lambda \alpha q)}
g)(\tau)=\int\!\!\!\!\!\!\!\!\!\!\!\!\!\!\int\limits_{\mathcal{D}_{(4)}(\lambda,
\tau, \alpha, q)} \!\!\!\!\!\!\!\!\!\! d\sigma dx\;
e^{-{(x/ 2)^2 \over T(\lambda)^2}}\;
X^\lambda_{-\lambda^{-2}\sigma+q-\left(\alpha+{1\over 2}\right)x}
A_{01}U_x A_{10} X^\lambda_{\lambda^{-2}\sigma-q+\left(\alpha-{1\over
2}\right)x} g(\sigma)
\end{equation}
where we have put $\mathcal{D}_{(4)}(\lambda, \tau, \alpha,
q))=\mathcal{D}_{(3)}(\lambda, \tau, \alpha, q)$. It turns out
that proving an operator convergence to
$\mathcal{H}^{(3)}_{(\lambda \alpha q)}$ is impossible, due to the
strong requirement of uniform convergence with respect to
\emph{any} normalized $g\in\mathcal{V}$. However, the Volterra
operators $\mathcal{H}^{(3)}_{(\lambda \alpha q)} $ and
$\mathcal{H}^{(4)}_{(\lambda \alpha q)}$ do fulfill the hypotheses
of Lemma \ref{lemma:mainest}, and so we conclude that
\begin{equation}
\lim_{\lambda\rightarrow 0} \|f^{(3)}_{(\lambda \alpha
q)}-f^{(4)}_{(\lambda \alpha q)}\|=0 \;.
\end{equation}

We now go back to consider the domain $\mathcal{D}_{(4)}(\lambda,
\tau, \alpha, q)$ as a function of $\lambda$, and note that it
tends to fill the strip
\begin{equation}
\mathcal{D}_{(5)}(\tau)= [0,\tau]\times[0,\infty] \;.
\end{equation}
Accordingly, we define the following Volterra integral operator on
this strip:
\begin{equation}
\mathcal{H}^{(5)}_{(\lambda \alpha q)} g= \!\!\!\int_0^\tau
\!\!\!d\sigma \!\int_0^{\infty}\!\!\!\!\!\! dx\,  e^{-{(x/2)^2
\over T(\lambda)^2}}\,
X^\lambda_{-\lambda^{-2}\sigma+q-\left(\alpha+{1\over 2}\right)x}
A_{01}U^\lambda_{x} A_{10} X^\lambda_{\lambda^{-2}\sigma-q+\left(\alpha-{1\over
2}\right)x} g(\sigma)\;.
\end{equation}
and note that
\begin{eqnarray}
&& \left\|\left(\left(\mathcal{H}^{(5)}_{(\lambda \alpha
q)}-\mathcal{H}^{(4)}_{(\lambda \alpha q)}\right)g\right)(\tau)
\right\|  \leq
\!\!\!\!\!\!\!\!\!\!\!\!\!\!\!\!\!\!\!\!\!\!\!\!\!
\iint\limits_{\quad\qquad\qquad\mathcal{D}^{(5)}(\tau)\setminus
\mathcal{D}^{(4)}(\lambda, \tau, \alpha, q)}
\!\!\!\!\!\!\!\!\!\!\!\!\!\!\!\!\!\!\!\!\!\!\!\!\!\!\!\!\!\!\!\!
d\sigma dx\;  \| A_{01}U_{x} A_{10} \| \; \|g\|_\infty \nonumber \\
&& \leq \|g\|_\infty \left\{
\lambda^2 q \int_{0}^{\infty} dx \;  \| A_{01}U_{x} A_{10} \| +\int_{\lambda^2 q}^\tau d\sigma\;
\int_{\overline{x}_l(\lambda,q,\sigma)}^{\infty} dx \;  \|
A_{01}U_{x} A_{10} \| \right\}
\end{eqnarray}
where
\begin{equation}
\overline{x}_l(\lambda,q,\sigma)={q-\lambda^{-2}\sigma \over
\alpha-{1\over 2}}
\end{equation}
is the $x$-coordinate of the left edge of the triangular domain
$\mathcal{D}(\lambda, \tau, \alpha, q)$ (see figure
(\ref{domain1})). Now the first term in the curly brackets clearly
goes to zero as $\lambda\rightarrow 0$ (with velocity $\sim
\lambda^2 q$), due to hypothesis (\ref{eq:no_oscillation_lambda})
for $\lambda=0$. In the second one, as before, we change
coordinate according to $\sigma\rightarrow\sigma-\lambda^2 q$,
obtaining
\begin{equation}
\int_0^{\tau-\lambda^2 q} \!\!\! d\sigma \int_{\overline{x}_l(\lambda,0,\sigma)}^{\infty}\!\!\! dx \,  \| A_{01}U_{x} A_{10} \| \;\leq\; \int_{0}^\tau d\sigma\;
\int_{\overline{x}_l(\lambda,0,\sigma)}^{\infty}\!\!\! dx \,  \|
A_{01}U_{x} A_{10} \|.
\end{equation}
This last term can be seen to go to zero uniformly on
$0\leq\tau\leq\overline{\tau}$ (and real $q$) by invoking, as done
before, the dominated convergence theorem, and using the fact that
$\overline{x}_l(\lambda,0,\sigma)>0$ a.e. in the (compact)
$\sigma$-integration domain (everywhere except $\sigma=0$). So it
follows that
\begin{equation}\label{eq:estimation_5}
\left\|\mathcal{H}^{(4)}_{(\lambda \alpha
q)}-\mathcal{H}^{(5)}_{(\lambda \alpha q)} \right\| \rightarrow 0,
\quad \lambda\rightarrow 0.
\end{equation}
Now we can finally compare with $\mathcal{H}^{(6)}_{(\lambda
\alpha q)}=\overline{\mathcal{H}}_{(\lambda\alpha q)}$: by adding
and subtracting obvious terms, we estimate
\begin{eqnarray}
&& \left\|\left(\mathcal{H}^{(5)}_{(\lambda \alpha q)}-\overline{\mathcal{H}}_{(\lambda \alpha q)}\right)g \right\| \nonumber\\
&&\leq \overline{\tau} \int_0^{\infty} dx\; \|X^\lambda_{q-\left(\alpha+{1\over 2}\right)x}-P_0 U_{q-\left(\alpha+{1\over 2}\right)x}\| \; \| A_{01}U_{x} A_{10} \|  \|g\|_\infty \nonumber \\
&&+ \overline{\tau} \int_0^{\infty} dx\; \| A_{01}U_{x} A_{10} \| \|X^\lambda_{\left(\alpha-{1\over 2}\right)x-q}-P_0 U_{\left(\alpha-{1\over 2}\right)x-q}\| \|g\|_\infty\;.
\end{eqnarray}
Again, uniform convergence to zero follows by hypothesis
(\ref{eq:no_oscillation_lambda}) together with the dominated
convergence theorem, using the fact that for every
$x\in\mathbb{R}$
\begin{equation}
\lim_{\lambda\rightarrow 0} \|X^\lambda_x-P_0 U_x\|=0.
\end{equation}
This proves the estimation in (\ref{eq:proof_1_conclusion}) and
thus finishes the proof.
$\quad\Box$ \\

\subsection{Comments to the theorem}
\begin{itemize}
\item The results of the theorem clearly generalize \cite{davies2}, as the semigroup studied there
is generated by our particular choice
$K_D=K_{(\alpha=1/2,q=0,T(\lambda)=+\infty)}$. It would in fact be
possible to prove the validity of Theorem \ref{th:K_(alpha_q_T)}
also for the case $T(\lambda)=+\infty$ identically, if the
particular choices $\alpha=\pm 1/2, q=0$ are made.

\item For each choice of $\alpha, q$ and $T$, the corresponding
generator $Z_0+\lambda A_{00}+K_{(\alpha,q,T(\lambda))}$ is always
well defined for all $\lambda \in [-1,1]$, $\lambda \neq 0$, no
matter which are $\mathcal{B}_0$ dimensions or $Z_0$'s spectral
properties.

\item The function $T:[-1,1]\rightarrow \mathbb{R}^+$ can be
considered as a measure of the system's transition times. Indeed,
an obvious and natural choice would be to put
\begin{equation}
T(\lambda)={1\over |\lambda|\:\|A\|}.
\end{equation}
With this choice, it is interesting to note that if we chose
$\tau=\|A\|^{-1}$ in the thesis of Theorem \ref{th:K_(alpha_q_T)},
the latter can be rewritten as
\begin{equation}
\lim_{\lambda\rightarrow 0} \left\{ \sup_{0\leq t\leq
\lambda^{-1}T(\lambda)}
\|W^\lambda_t-\overline{W}^\lambda_t\|\right\}=0,
\end{equation}
so that we see that the approximation is valid up to the
"dynamical observation time" $\lambda^{-1}T(\lambda)$, which is greater than
the "transition time" $T(\lambda)$, but shorter than possible
Poincar{\'e}-like recurrence times, as
$\lambda^{-1}T(\lambda)<+\infty$ for any nonzero $\lambda$. This
scaling transition time $T(\lambda)$ will play an important role
in our second main theorem, through the definition of a dynamical
time average.

\item Even if the generator in Theorem \ref{th:K_(alpha_q_T)} is
always well defined for nonzero values of the coupling constant
$\lambda$, it does not give rise to a Contraction Semigroup, as can be easily seen. This means that we still have to perform some
kind of temporal average, like the one introduced in
\cite{davies1}, in order to obtain a contractive
semigroup. We shall do it in the next section.
\end{itemize}

\subsection{A Sufficient Condition For the Hypotheses}
We provide a sufficient condition for the validity of the
hypotheses in Theorem \ref{th:K_(alpha_q_T)}, which is a slight
adaptation of Theorem 1.3 in \cite{davies2}, and is physically supported by perturbation argumentations.
First, define
\begin{equation}
a_n(t)=\int_0^t dt_0 \cdots \int_0^{t_{n-1}} dt_n
\|A_{01}U_{t_0-t_1} A_{11} U_{t_1-t_2} A_{11}\ldots A_{11}
U_{t_{n}}A_{10} \|,
\end{equation}
which are, as one can easily realize, coming from the expansion
coefficients of $U^\lambda_t$ within the subspace $\mathcal{B}_1$
in powers of $\lambda$:
\begin{equation}
\int_0^t \|A_{01}U^\lambda_{t_0} A_{10}\| \, dt_0 \leq \sum_{n=0}^\infty \lambda^n a_n(t).
\end{equation}
\begin{theorem}
Suppose that
\begin{equation}\label{hp:infrectime}
\int_0^\infty \|A_{01}U_{t_0} A_{10}\| \, dt_0 < \infty.
\end{equation}
Suppose that
\begin{equation}
a_n(t)\leq c_n |t|^{n/2}
\end{equation}
for all $t\in\mathbb{R}$ and $n\geq 1$, where the series
$\sum_{n=1}^\infty c_n z^n$ has infinite radius of convergence.
Suppose also that for some $\epsilon>0$, $d_n$, and all $t\geq 0$
\begin{equation}
a_n(t)\leq d_n |t|^{n/2-\epsilon}
\end{equation}
Then the conditions of Theorem \ref{th:K_(alpha_q_T)} are
satisfied, namely, there exists some $0<c<\infty$ such that for
every $\overline{\tau}>0$
\begin{equation}
\int_0^{\lambda^{-2}\overline{\tau}} \|A_{01}U^\lambda_x A_{10}\|
\, dx \leq c
\end{equation}
uniformly on $|\lambda|\leq 1$, and also, for every
$0<\overline{\tau}<\infty$,
\begin{equation}
\lim_{\lambda\rightarrow 0} \int_0^{\lambda^{-2}\overline{\tau}}
\|A_{01}(U^\lambda_x - U_x) A_{10}\| \, dx = 0 \;.
\end{equation}
\end{theorem}

{\bf Proof}.
By expanding $U^\lambda_x$ in a $\lambda$ power series, one
obtains
\begin{equation}
\int_0^{\lambda^{-2}\overline{\tau}} \|A_{01}U^\lambda_x A_{10}\|
\, dx \leq \sum_{n=0}^\infty\lambda^n a_n(\lambda^{-2}\overline{\tau})
\leq \sum_{n=0}^\infty c_n |\overline{\tau}|^{n/2}
\end{equation}
which converges for any $\overline{\tau}$. Similarly,
\begin{equation}
\int_0^{\lambda^{-2}\overline{\tau}} \|A_{01}(U^\lambda_x - U_x)
A_{10}\| \, dx \leq \sum_{n=1}^\infty \lambda^n
a_n(\lambda^{-2}\overline{\tau})
\leq \sum_{n=1}^\infty d_n |\lambda|^{2\epsilon}
|\overline{\tau}|^{n/2}.
\end{equation}
For $|\lambda|<1$ the series is dominated by the convergent
$\sum_{n=1}^\infty c_n |\overline{\tau}|^{n/2}$, and each term of
the series goes to zero when $\lambda\rightarrow 0$, completing
the proof.
$\quad\Box$ \\
Clearly, these conditions refer to the decay properties of the $n$-point correlation functions $a_n(t)$, and show that the hypotheses (\ref{eq:no_oscillation_lambda}) and (\ref{hp:lambda_convergence}) are satisfied provided information flows fast enough from the subsystem $\mathcal{B}_0$ to the remaining degrees of freedom $\mathcal{B}_1$. It is important to note that these conditions however only refer to the possibility to perform a semigroup approximation, thereby eliminating in markovian fashion the memory kernel, and thus only account for irreversibility: they are not directly linked to obtaining a dissipative process, or at least, they are not sufficient. To account for dissipation, we need the results of the next two sections.

\section{Dynamical Time Averaging Map}
Our second main result will be to pick, among the possible choices
just found for the semigroup generator, the most symmetric, and
perform a "dynamical" temporal average, which will always be well
defined, no matter which are the spectral properties of $Z_0$ or
the dimensions of $\mathcal{B}_0$. The term "dynamical" here means
that we shall find an alternative averaging map, different from
the one introduced in \cite{davies1}, that depends on the coupling
constant, to remove the singularities that so severely limit the
usual time average. The final and remarkably symmetric form of the
dynamically averaged operator will then be shown to be again
compatible with the exact evolution. But what is most important,
it will be shown to generate a Contraction Semigroup in the
next section.

In \cite{davies1} a spectral averaging is introduced: for an
operator $K: \mathcal{B}_0 \rightarrow \mathcal{B}_0$, we put
\begin{equation}\label{eq:spavdav}
K^\natural=\lim_{T\rightarrow +\infty} {1\over 2T}\int_{-T}^T dq\;
U_{-q} K U_q
\end{equation}
whenever the right hand side is defined. It is important to note here that one has to take the limit $T\rightarrow +\infty$ in order to perform the required time averaging (this in fact allows to diagonalize $K$ in the finite dimensional case). Then Davies shows in
\cite{davies2} that if $\mathcal{B}_0$ is finite dimensional, then
the operation $\natural$ is well defined and for every
$\overline{\tau}>0$
\begin{equation}
\lim_{\lambda\rightarrow 0}\left\{ \sup_{0\leq t\leq
\lambda^{-2}\overline{\tau}} \left\|e^{(Z_0+\lambda^2 K)
t}-e^{(Z_0+\lambda^2 K^\natural) t} \right\| \right\} =0.
\end{equation}
In the important example studied in \cite{davies1}, the author
finds a completely positive dynamics~\cite{lindblad} for a finite dimensional
system coupled to a heat bath using the operator $K_D^\natural$,
where $K_D$ is defined in (\ref{def:K_D}). However, in \cite{davies2} the author
shows that the temporal average introduce in \cite{davies1} "was something of a red herring".
On the other hand the evolution generated by the unaveraged $K_D$ is no more positive in general, so the role of
the time average becomes instead somewhat important for
positivity. Unfortunately, as said before, the time average
$\natural$ is not generally defined when the system hamiltonian
has continuous spectrum. Since in this work we would like to remain in the contest of a generic Banach space, we won't study positivity, but we shall focus on more general dissipative properties of the generators.

Thanks to Theorem \ref{th:K_(alpha_q_T)}, we are now
ready to introduce a new type of temporal averaging, that will scale with the coupling constant $\lambda$ and will
always be well defined (except possibly the singular and uninteresting case $\lambda=0$).

We start choosing $\alpha=0$, which seems a rather symmetrical case. Now
let's fix some positive T, and note that
\begin{eqnarray}
K_{(0,q,T)} &=& U_q \left\{\! \int_0^\infty \!\!\! dx\,
e^{-\left({x\over 2}\right)^2 / T^2}\,
U_{-{x\over 2}} A_{01} U_x A_{10} U_{-{x\over 2}} \right\} U_{-q} \nonumber \\
&=& U_q K_{0,0,T} U_{-q}.
\end{eqnarray}

The idea is now the following: since in the limit of small coupling
$\lambda\rightarrow 0$ the system transition time goes to
$T(\lambda)\rightarrow +\infty$, we could use it to diagonalise the generator $K_{(0,0,T(\lambda))}$ by just performing a
$T(\lambda)$-dependent gaussian integration in the $q$-variable. This would
allow the time average of $K_{(0,0,T(\lambda))}$ to depend on $\lambda$ (hence the name "dynamical"), and
to be well defined everytime $\lambda\neq 0$. To this purpose we give the following
\begin{definition}\label{def:K_T}
For any real positive $T>0$ put
\begin{eqnarray}
K_T &=& {1\over \sqrt{\pi}T} \int_{-\infty}^\infty \!\!\! dq\, e^{-q^2
/ T^2} K_{(0,q,T)} \nonumber\\
&=& {1\over \sqrt{\pi}T}
\int_{-\infty}^\infty \!\!\! dq\, e^{-q^2/ T^2} \int_0^\infty
\!\!\! dx\, e^{-\left({x\over 2}\right)^2 / T^2}\,
U_{-{x\over 2}+q} A_{01} U_x A_{10} U_{-{x\over 2}-q}  \nonumber \\
&=& {1\over \sqrt{\pi}T} \int_{-\infty}^\infty  dt_1\, e^{-{t_1^2
\over 2 T^2}}\; A_{01}(t_1) \int_{-\infty}^{t_1} dt_2\, e^{-{t_2^2
\over 2 T^2}}\; A_{10}(t_2)
\end{eqnarray}
where we have denoted $A_{ij}(t)=U_{-t} A_{ij} U_{t}$.
\end{definition}
Note that in the last line we have changed variable according to
\begin{equation}
\left\{
\begin{array}{c}
t_1=x/2-q \\
t_2=-x/2-q
\end{array}
\right.
\end{equation}
In passing, we observe that $K_T$ so defined can be easily written in more physical terms as
\begin{equation}
K_T = {1\over 2\sqrt{\pi}T} \int_{-\infty}^{+\infty}  dt_1 \int_{-\infty}^{+\infty} dt_2\; e^{-{t_1^2+t_2^2
\over 2 T^2}}\; \mathcal{T}[A_{01} A_{10}](t_1,t_2)
\end{equation}
where we have introduced the time ordering
\begin{equation}
\mathcal{T}[A_{01} A_{10}](t_1,t_2)=A_{01}(t_1) A_{10}(t_2)\theta(t_1-t_2)+A_{01}(t_2) A_{10}(t_1)\theta(t_2-t_1),
\end{equation}
$\theta$ being the Heaviside step function. Indeed, this expression closely resembles well known von Neumann and Dyson series expansion for the unitary evolution operator of a closed quantum system (see for example \cite{fetter_walecka}), and to our opinion could offer a great help in understanding how the Nakajima-Prigogine-Resibois-Zwanzig master equation (\ref{eq:exact}) could be correctly approximated beyond second order\footnote{Another important explicit form for $K_T$ will be addressed in the proof of the next section}.

\begin{theorem}\label{th:K_T}
Suppose that $X^\lambda_t$ is a one-parameter group of isometries.
Suppose that there exists some $0<c<\infty$ such that for every
$\overline{\tau}>0$
\begin{equation}\label{eq:no_oscillation_lambda_second}
\int_0^{\lambda^{-2}\overline{\tau}} \|A_{01}U^\lambda_x A_{10}\|
\, dx \leq c
\end{equation}
uniformly on $|\lambda|\leq 1$. Suppose also that for every
$0<\overline{\tau}<\infty$
\begin{equation}\label{hp:lambda_convergence_second}
\lim_{\lambda\rightarrow 0} \int_0^{\lambda^{-2}\overline{\tau}}
\|A_{01}(U^\lambda_x - U_x) A_{10}\| \, dx = 0 \;.
\end{equation}
Let $T\in\mathcal{C}([-1,1],\overline{\mathbb{R}})$ be a real valued positive
continuous function on the interval $[-1,1]$, such that
\begin{equation}
T(\lambda)\sim |\lambda|^{-\xi} \widetilde{T} ,\quad
\lambda\sim 0
\end{equation}
for some real positive reference time $\widetilde{T}>0$ and
scaling $0<\xi<2$. Denote with
\begin{equation}\label{def:semigr_T}
\widetilde{W}^\lambda_t=\exp\{(Z_0+\lambda A_{00}+\lambda^2
K_{T(\lambda)})t\}
\end{equation}
the associated semigroup on $\mathcal{B}_0$.

Then for every $\overline{\tau}>0$
\begin{equation}
\lim_{\lambda\rightarrow 0} \left\{ \sup_{0\leq t\leq
\lambda^{-2}\overline{\tau}}
\|W^\lambda_t-\widetilde{W}^\lambda_t\|\right\}=0.
\end{equation}
\end{theorem}

{\bf Proof}.
We shall borrow most part of the proof of Theorem
(\ref{th:K_(alpha_q_T)}). Accordingly, we should denote for
example with $\mathcal{H}_{(\lambda q)}$ the operator defined in
Theorem (\ref{th:K_(alpha_q_T)}) as $\mathcal{H}_{(\lambda 0 q)}$,
and so on.

Define
\begin{equation}
\mathcal{H}_\lambda^{(j)}={1\over \sqrt{\pi} T(\lambda)}
\int_{-\infty}^\infty \!\!\! dq\, e^{-q^2 / T(\lambda)^2}
\;\mathcal{H}_{(\lambda q)}^{(j)} \;.
\end{equation}
This is obviously Volterra, being an integral of Volterra
operators. A closer inspection soon reveals that also
\begin{equation}
\|\mathcal{H}_\lambda^{(j)}\| \leq {1\over \sqrt{\pi} T(\lambda)}
\int_{-\infty}^\infty \!\!\! dq\, e^{-q^2 / T(\lambda)^2}
\;\|\mathcal{H}_{(\lambda q)}^{(j)}\| \leq c\overline{\tau}\;,
\end{equation}
as for each $j$, $\mathcal{H}_{(\lambda q)}^{(j)}$ is bounded by
$c$ uniformly on $q$ and
\begin{equation}\label{eq:normalization}
{1\over \sqrt{\pi} T(\lambda)} \int_{-\infty}^\infty  dq\, e^{-q^2
/ T(\lambda)^2} =1 \;.
\end{equation}
We proceed on the very same lines of Theorem
(\ref{th:K_(alpha_q_T)}): the proof that
\begin{equation}
\lim_{\lambda\rightarrow 0}
\|\mathcal{H}^{(j)}_\lambda-\mathcal{H}_\lambda^{(j-1)}\|=0
\end{equation}
for $j=1,2$, and $j=6$ are in fact identical to that of the
foretold Theorem, as, for those values for $j$, one has
\begin{equation}\label{eq:estimation_coarse}
\lim_{\lambda\rightarrow 0}
\|\mathcal{H}_{(\lambda,q)}^{(j)}-\mathcal{H}_{(\lambda,q)}^{(j-1)}\|=0
\end{equation}
uniformly on $q$. Property (\ref{eq:normalization}) can be exploited to state that
\begin{equation}
\lim_{\lambda\rightarrow 0}
\|\mathcal{H}_\lambda^{(j)}-\mathcal{H}_\lambda^{(j-1)}\|=0
\end{equation}
for $j=3$ and $j=5$. In fact, for these values of $j$, we can
estimate
\begin{equation}
\|\mathcal{H}_\lambda^{(j)}-\mathcal{H}_\lambda^{(j-1)}\| \leq
{1\over \sqrt{\pi} T(\lambda)} \int_{-\infty}^\infty  dq\, e^{-q^2
/ T(\lambda)^2} \;
\|\mathcal{H}_{(\lambda,q)}^{(j)}-\mathcal{H}_{(\lambda,q)}^{(j-1)}\|
\;.
\end{equation}
Now, the norm in the integrand goes to zero as $\sim
c^{(j)}_1(\lambda)+c^{(j)}_2 \lambda^2 q$, with
$c^{(j)}_1(\lambda)\rightarrow 0$ uniformly on $q$ as
$\lambda\rightarrow 0$, as already noted in the proof of Theorem
(\ref{th:K_(alpha_q_T)}), so the whole integral goes to zero as
$\lambda\rightarrow 0$ precisely because $T(\lambda)$ scales with
$|\lambda|^{-\xi}$ and $\xi<2$ by our hypothesis.

It remains to show that if $f_\lambda^{(j)}=\sum_n
\left(\mathcal{H}_\lambda^{(j)}\right)^n b$, for some initial condition $b\in\mathcal{B}_0$,
then
\begin{equation}\label{eq:estimation_qq'}
\lim_{\lambda\rightarrow 0} \|f_\lambda^{(4)}-f_\lambda^{(3)}\|=0
\;.
\end{equation}
Proceeding according to Lemma \ref{lemma:mainest}, we
estimate
\begin{eqnarray}\label{eq:est1qq'}
\|f_\lambda^{(3)}-f_\lambda^{(4)}\|_\infty
&\leq& \sum_{n=1}^\infty\; \sum_{l=1}^{n-1}\; {(\overline{\tau}\:c)^{n-l-1} \over (n-l)!} \; {1\over \sqrt{\pi} T(\lambda)} \int_{-\infty}^\infty  dq\, e^{-q^2 / T(\lambda)^2} \nonumber\\
&&\times \left\|(\mathcal{H}_{(\lambda q)}^{(3)}-\mathcal{H}_{(\lambda
q)}^{(4)})\:\left(\mathcal{H}_{\lambda}^{(4)}\right)^l\:
b\right\|_\infty \;.
\end{eqnarray}
Now we compute
\begin{eqnarray}
&& \left[\left(\mathcal{H}_{(\lambda q)}^{(3)}-\mathcal{H}_{(\lambda q)}^{(4)}\right)\:\left(\mathcal{H}_{\lambda}^{(4)}\right)^l\: b\right](\tau)\nonumber \\
&=& \iint\limits_{\overline{\mathcal{D}}(\lambda, \tau, q)} \!\!\!\! d\sigma dx \, e^{-{(x/ 2)^2\over T(\lambda)^2}}\, K_q(\lambda,\sigma,x)  \left\{ \left[\left(\mathcal{H}_{\lambda}^{(4)}\right)^l b\right] (\sigma\!-\!\lambda^2 (q \! -\!x/2))\!-\!\left[\left(\mathcal{H}_{\lambda}^{(4)}\right)^l b\right](\sigma) \right\} \nonumber \\
&=& {1\over \sqrt{\pi} T(\lambda)} \int_{-\infty}^\infty  dq'\, e^{-q'^2 / T(\lambda)^2} \iint\limits_{\overline{\mathcal{D}}(\lambda, \tau, q)} \!\!\!\! d\sigma dx \, e^{-{(x/ 2)^2\over T(\lambda)^2}}\; K_q(\lambda,\sigma,x) \nonumber\\
&&\times \left\{ \left[\mathcal{H}_{\lambda q'}^{(4)}\left(\left(\mathcal{H}_{\lambda}^{(4)}\right)^{l-1} b\right)\right]\left(\sigma-\lambda^2 (q -x/2)\right) -\left[\mathcal{H}_{\lambda q'}^{(4)}\left(\left(\mathcal{H}_{\lambda}^{(4)}\right)^{l-1} b\right)\right](\sigma) \right\} \nonumber \\
&=& {1\over \sqrt{\pi} T(\lambda)} \int_{-\infty}^\infty  dq'\, e^{-q'^2 / T(\lambda)^2} \iint\limits_{\overline{\mathcal{D}}(\lambda, \tau, q)}\!\!\!\! d\sigma dx \, e^{-{(x/ 2)^2\over T(\lambda)^2}}\; K_q(\lambda,\sigma,x) \nonumber\\
&&\times\iint\limits_{\widetilde{S}(\lambda,q,q',\sigma,x)}
d\sigma_1 dx_1 \, \Delta^\lambda(\sigma_1,x_1)\,e^{-{(x/ 2)^2\over
T(\lambda)^2}}\; K_{q'}(\lambda,\sigma_1,x_1)\,
\left[\left(\mathcal{H}_{\lambda}^{(4)}\right)^{l-1}
b\right](\sigma_1)
\end{eqnarray}
where we have put $\Delta^\lambda=\chi^\lambda_1-\chi^\lambda_2$,
$\chi^\lambda_1$ being the characteristic function of
$\overline{\mathcal{D}}(\lambda, \sigma-\lambda^2 \left(q
-x/2\right), q')$,  $\chi^\lambda_2$ the characteristic function
of $\overline{\mathcal{D}}(\lambda, \sigma, q')$, and we have
defined
\begin{equation}
\widetilde{S}(\lambda,q,q',\sigma,x)=\overline{\mathcal{D}}(\lambda,
\sigma-\lambda^2 \left(q -x/2\right), q')\cup
\overline{\mathcal{D}}(\lambda, \sigma, q')\;.
\end{equation}
Passing to the norms we obtain
\begin{eqnarray}\label{eq:est2qq'}
&&\|(\mathcal{H}_{(\lambda q)}^{(3)}-\mathcal{H}_{(\lambda q)}^{(4)})\:{\mathcal{H}_{\lambda}^{(4)}}^l\: b\|_\infty \leq {(\overline{\tau}\:c)^{l-1} \over (l-1)!} \:\|b\|_{\mathcal{B}_0} {1\over \sqrt{\pi} T(\lambda)} \int_{-\infty}^\infty  dq'\, e^{-q'^2 / T(\lambda)^2} \nonumber\\
&&\quad\times \sup_{0<\tau<\overline{\tau}} \quad \iint\limits_{\mathcal{D}(\lambda, \tau, q)} \!\! d\sigma dx \; e^{-(x/2)^2 / T(\lambda)^2} k(x) \; \Xi_{q q'}(\lambda,\sigma,x)
\end{eqnarray}
where the slight modification of the related definition of
$\Xi_{q}(\lambda,\sigma,x)$ in (\ref{def:Xi}) is given by
\begin{equation}\label{def:Xiqq}
\Xi_{q q'}(\lambda,\sigma,x)
=\iint\limits_{\widetilde{S}(\lambda,q,q',\sigma,x)}
\!\!\!d\sigma_1 dx_1 \, |\Delta^\lambda(\sigma_1,x_1)|\,k(x_1) \;.
\end{equation}
Proceeding on the same lines as in Lemma \ref{lemma:mainest} we
find the asymptotic behavior
\begin{eqnarray}\label{eq:asymptqq'}
&&\sup_{0<\tau<\overline{\tau}} \quad \iint\limits_{\mathcal{D}(\lambda, \tau, q)} \!\! d\sigma dx \; e^{-(x/2)^2 / T(\lambda)^2} k(x) \; \Xi_{q q'}(\lambda,\sigma,x) \nonumber\\
&&\leq\overline{\tau} \sup_{0\leq\sigma\leq\overline{\tau}} \int_0^\infty dx e^{-(x/2)^2 / T(\lambda)^2} k(x) \; \Xi_{q q'}(\lambda,\sigma,x) \nonumber\\
&&\leq \overline{\tau} \{ c_1(\lambda)+c_2 \lambda^2 |q| + c_3
\lambda^2 |q'| \}
\end{eqnarray}
where $c_1(\lambda)\rightarrow 0$ uniformly on $q$ and $q'$. This,
plus the fact that $\xi<2$ by our hypothesis, allows us conclude
that Eq. (\ref{eq:estimation_qq'}) holds, as can be seen by
putting result (\ref{eq:asymptqq'}) into (\ref{eq:est2qq'}), and
then back into (\ref{eq:est1qq'}), and by using the dominated
convergence theorem.

Collecting the results as in Theorem (\ref{th:K_(alpha_q_T)})
concludes the proof.
$\quad\Box$ \\

To make contact with the definition of the time average proposed
in \cite{davies1}, we give the following
\begin{proposition}
Let $\mathcal{B}_0$ be finite dimensional, and $A_{00}=0$. Then for $b\in\mathcal{B}_0$ and every $\tau>0$
\begin{equation}
\lim_{\lambda\rightarrow 0} \left\{\sup_{0\leq t\leq \lambda^{-2}\tau} \|e^{(Z_0+\lambda^2 K_{T(\lambda)})t}b-e^{(Z_0+\lambda^2 K_D^\natural)t}b\|\right\} =0
\end{equation}
\end{proposition}
{\bf Proof}.
Let
\begin{equation}
Z_0=\sum_\alpha i \omega_\alpha Q_\alpha
\end{equation}
be the spectral decomposition of $Z_0$, with all $\omega_\alpha$'s distinct and real, and compute
\begin{eqnarray}
\lim_{T\rightarrow +\infty} K_T &=& \lim_{T\rightarrow +\infty} {1\over \sqrt\pi T} \int_{-\infty}^\infty dq\; e^{-q^2/T^2}\sum_{\alpha\beta} e^{i(\omega_\alpha-\omega_\beta)q} Q_\alpha K_{(0,0,+\infty)} Q_\beta \nonumber\\
&=&\sum_\alpha Q_\alpha K_D Q_\alpha \nonumber\\
&=& \lim_{T\rightarrow +\infty} {1\over 2 T}\int_{-T}^T dq\; \sum_{\alpha\beta} e^{i(\omega_\alpha-\omega_\beta)q}\,Q_\alpha K_D Q_\beta \nonumber \\
&=& \lim_{T\rightarrow +\infty} {1\over 2T}\int_{-T}^T dq\;
U_{-q} K_D U_q
\end{eqnarray}
which clearly shows that the time average in \cite{davies1}
coincides with our dynamical one, in the
weak-coupling limit $\lambda\rightarrow 0$, that is, recalling that $\lim_{\lambda\rightarrow 0} T(\lambda)= +\infty$,
\begin{equation}
K^\natural_D= \lim_{\lambda\rightarrow 0} K_{T(\lambda)}
\end{equation}
and the averaging map $\natural$ is well defined because $\mathcal{B}_0$ is finite dimensional. Then a slight and straightforward modification of Theorem 1.4 in~\cite{davies2} proves the stated result.
$\quad\Box$ \\

\begin{itemize}
\item We stress that, contrary to $K_D^\natural$, our time average $K_{T(\lambda)}$ is always well defined, irrespective of $Z_0$
spectral properties, for any nonzero value of the coupling constant $\lambda$.

\item The statement of the proposition would remain unchanged if we relax the hypothesis that $\mathcal{B}_0$ is finite dimensional, and only assume that $Z_0$ has discrete spectrum.
\end{itemize}

\subsection{Comments To The Theorem}
The last line in Definition \ref{def:K_T} indicates a high level
of symmetry and simplicity, so that one could ask whether such a form would be "unique" to some
extent. With respect to this, we remark here that there is nothing
peculiar in the choice $\alpha=0$ that led us to the dynamical
time average in Definition \ref{def:K_T}, apart from simplicity in
the definitions and in the proofs involved. In fact, one could
equally well proceed along the following lines. In equation
(\ref{eq:exact}) change variable according to (\ref{eq:vartr_RT})
as in Theorem \ref{th:K_(alpha_q_T)}. Proceed along the lines of
Theorem \ref{th:K_(alpha_q_T)}, but use
\begin{equation}
\exp\{-t_1(x,q)^2+t_2(x,q)^2-2q^2/ 2T(\lambda)^2\}
\end{equation}
instead of $\exp\{-(x/2)^2/T(\lambda)^2\}$ as gaussian smoothing
for the kernel in (\ref{def:h2}), where
\begin{equation}
\left\{
\begin{array}{c}
t_1(x,q)=(\alpha+1/2)x-q \\
t_2(x,q)=(\alpha-1/2)x-q
\end{array}
\right. \;.
\end{equation}
There is no difficulty to proceed as in Theorem
\ref{th:K_(alpha_q_T)} to show that (under the same hypotheses)
the semigroup $\widehat{W}_t^\lambda=\exp\{(Z_0+\lambda
A_{00}+\lambda^2 \widehat{K}_{(\alpha,q,T(\lambda))})t\}$
satisfies
\begin{equation}
\lim_{\lambda\rightarrow 0} \left\{ \sup_{0\leq t\leq
\lambda^{-2}\overline{\tau}}
\|W^\lambda_t-\widehat{W}^\lambda_t\|\right\}=0,
\end{equation}
for every $\overline{\tau}>0$, with the modified version
\begin{eqnarray}
\widehat{K}_{(\alpha,q,T(\lambda))}&=&\int_0^\infty dx\;
e^{-{t_1(x,q)^2+t_2(x,q)^2-2q^2\over 2T(\lambda)^2}}\;
U_{-\left(\alpha+{1\over 2}\right)x+q} A_{01} U_x A_{10}
U_{\left(\alpha-{1\over 2}\right)x-q}\nonumber\\
&=& \int_0^\infty dx\;
e^{-{t_1(x,q)^2+t_2(x,q)^2-2q^2\over 2T(\lambda)^2}}\; A_{01}(t_1(x,q))
A_{10}(t_2(x,q))\;.
\end{eqnarray}
Then as in Theorem \ref{th:K_T}, one could perform a gaussian
integration on the $q$ variable to obtain
\begin{eqnarray}
K_{T(\lambda)}&=&{1\over \sqrt{\pi}T(\lambda)} \int_{-\infty}^\infty \!\!\!
dq\, e^{-q^2 / T(\lambda)^2}\;\widehat{K}_{(\alpha,q,T(\lambda))}
\nonumber\\
&=& {1\over \sqrt{\pi}T(\lambda)} \int_{-\infty}^\infty  dt_1\, e^{-{t_1^2
\over 2 T(\lambda)^2}}\; A_{01}(t_1) \int_{-\infty}^{t_1} dt_2\, e^{-{t_2^2
\over 2 T(\lambda)^2}}\; A_{10}(t_2)
\end{eqnarray}
exactly as in Definition \ref{def:K_T}. This shows that the choice
$\alpha=0$ is merely dictated by the simplicity of the proofs
involved, and furnishes an argument for the uniqueness of
expression \ref{def:K_T}. One could argue that the choice of a
gaussian to smooth the involved kernels is somewhat arbitrary,
but, as commented in \cite{tajpra}, the gaussian is the only
distribution $\Phi$ with the factorization property
\begin{equation}
\Phi(q)\Phi(x/2)=\Phi(t_1/\sqrt{2})\Phi(t_2/\sqrt{2})
\end{equation}
for $t_1=q+x/2$ and $t_2=q-x/2$. As the remaining degree of
arbitrariness is concerned, the transition time $T(\lambda)$, we have already observed that it is actually fixed
by the obvious and natural choice $T(\lambda)=\| \lambda A \|^{-1}$.

\section{A Contraction Semigroup}
To establish the extent to which $K_T$ is unique, we recall that $\mathcal{T}_t$ is a contraction semigroup if $\|\mathcal{T}_t\|\leq 1$ for all $t>0$.

Before stating our main result of this section, we report here for completeness the Hille-Yosida Theorem~\cite{hille}, an important part of which will be used throughout in our statements (although the operators involved will be bounded). We denote with $R(S)$ the range of an operator $S$.
\begin{theorem}[Hille-Yosida theorem]
Let $S$ be an operator on the Banach space $\mathcal{B}$. For $\mathcal{F}=\mathcal{B}_*$ or $\mathcal{F}=\mathcal{B}^*$ the following conditions are equivalent:
\begin{enumerate}
\item $S$ is the infinitesimal generator of a $\sigma(\mathcal{B},\mathcal{F})$-continuous semigroup of contractions $\mathcal{T}$;
\item $S$ is $\sigma(\mathcal{B},\mathcal{F})$-densely defined, and $\sigma(\mathcal{B},\mathcal{F})$-$\sigma(\mathcal{B},\mathcal{F})$-closed. For $\alpha>0$
    \begin{equation}\label{eq:hille_Yosida}
    \|(1-\alpha S)b\| \geq \|b\|, \qquad b\in D(S),
    \end{equation}
    and for some $\alpha>0$
    \begin{equation}
    R(1-\alpha S)=\mathcal{B}
    \end{equation}
\end{enumerate}
If these conditions are satisfied, then the semigroup defined in terms of $S$ by either of the limits
\begin{eqnarray}
\mathcal{T}_t b &=& \lim_{\epsilon\rightarrow 0}\exp\{tS(1-\epsilon S)^{-1}\} b \nonumber\\
&=& \lim_{n\rightarrow\infty} (1-S t/n)^n b
\end{eqnarray}
where the exponential of the bounded $tS(1-\epsilon S)^{-1}$ is defined by power series expansion. The limits exist in the $\sigma(\mathcal{B},\mathcal{F})$ topology, uniformly for $t$ in compacts, and in norm if $b\in\overline{D}(S)$ is in the norm closure of $D(S)$.
\end{theorem}
We shall also need the following:
\begin{lemma}\label{lemma:contraction}
Let $A$ and $B$ be bounded generators of one-parameter groups of isometries on a Banach space $\mathcal{B}$.

Then $C=[A,B]$ is the generator of a one-parameter group of isometries on $\mathcal{B}$.
\end{lemma}
{\bf Proof}.
Define
\begin{equation}
F_t=\exp\{At\} \exp\{Bt\}\exp\{-At\}\exp\{-Bt\} \qquad t\in\mathbb{R}
\end{equation}
Clearly $F_t$ is an isometry for every $t\in\mathbb{R}$. Now for $t\in\mathbb{R}\setminus\{0\}$ define
\begin{equation}
C_n(t)=n^2 t^{-2}(F_{t/n}-1).
\end{equation}
Then $C_n(t)$ generates a one-parameter group of isometries. To show this we take $\alpha\geq 0$, $b\in\mathcal{B}$, and prove inequality (\ref{eq:hille_Yosida}) by computing
\begin{eqnarray}
&&\|(1-\alpha(F_{t/n}-1))b\|=(1+\alpha)\|(1-{\alpha\over 1+\alpha} F_{t/n})b\| \nonumber\\
&&\;\;\geq (1+\alpha)\left|\|b\|-{\alpha\over 1+\alpha}\|F_{t/n}b\| \right| \; \geq \; \|b\|,
\end{eqnarray}
as $\|b\|\geq \alpha/ (1+\alpha)\:\|F_{t/n}b\|$ because $F_{t/n}$ is an isometry and $\alpha/ (1+\alpha)\:\leq\: 1$. It can readily be seen that all the other hypotheses of the Hille-Yosida Theorem are satisfied (recall that $A$ and $B$ are supposed to be bounded). But proceding in the same way, inequality (\ref{eq:hille_Yosida}) can be seen to hold also for $\alpha<0$, so that $C_n(T)$ actually generates a one-parameter group of isometries~\cite{bratteli_robinson}.

Now a simple calculation~\cite{goldstein} shows that $\lim_{n\rightarrow\infty} C_n(t) b=C b$ for every $b\in\mathcal{B}$. But the inequality (\ref{eq:hille_Yosida}), together with the validity of all the other conditions of the Hille-Yosida Theorem, passes to the limit, due to the boundedness of the involved operators, thus proving the Lemma.
$\quad\Box$ \\
We can now state the main result of this section:
\begin{theorem}
If $\|P_0\|=1$, then $\widetilde{W}^\lambda_t$ is a contraction semigroup on $\mathcal{B}_0$, for all real $\lambda$.
\end{theorem}
{\bf Proof}.
First of all $X^\lambda_t$ is a one-parameter group of isometries because of Lemma \ref{lemma:davies}. Now because of the Trotter product formula~\cite{trotter}, one has
\begin{equation}
P_0\widetilde{W}^\lambda_t=P_0 e^{(Z_0+\lambda A_{00}+\lambda^2 K_{T(\lambda)})t}=\lim_{n\rightarrow\infty} \left\{X^\lambda_{t/n}\: e^{\lambda^2 K_{T(\lambda)} {t/ n}}\right\}^n,
\end{equation}
so that
\begin{equation}
\|P_0\widetilde{W}^\lambda_t\|\leq \sup_n \| P_0 e^{\lambda^2 K_{T(\lambda)} {t/ n}}\|^n,
\end{equation}
and the theorem would follow if $K_T$ would generate a contraction semigroup on $\mathcal{B}_0$, for all $T>0$. To show this is indeed the case, we name
\begin{equation}
\Phi(t)=\sqrt{1\over \sqrt{\pi}T}\; e^{-{t^2\over 2T^2}}\; U_{-t}A U_t
\end{equation}
and denote as usual $\Phi_{ij}(t)=P_i\Phi(t) P_j$. Then from
\begin{eqnarray}
K_T &=& \int_{-\infty}^{+\infty} \!\!\! dt_1 \int_{-\infty}^{+\infty}\!\!\! dt_2\; \Phi_{01}(t_1) \Phi_{10}(t_2) -\int_{-\infty}^{+\infty} \!\!\! dt_1\int_{-\infty}^{t_1} \!\!\! dt_2\;  \Phi_{01}(t_2) \Phi_{10}(t_1) \nonumber\\
K_T &=&  \int_{-\infty}^{+\infty} \!\!\! dt_1 \int_{-\infty}^{t_1}\!\!\! dt_2 \; \Phi_{01}(t_1) \Phi_{10}(t_2)
\end{eqnarray}
we sum term by term to obtain
\begin{equation}
K_T = {1\over 2} \int_{-\infty}^{+\infty} \!\!\! dt_1 \; \Phi_{01}(t_1) \int_{-\infty}^{+\infty} \!\!\! dt_2 \; \Phi_{10}(t_2) + {1\over 2} \int_{-\infty}^{+\infty} \!\!\! dt_1 \int_{-\infty}^{t_1} \!\!\! dt_2\;  [\Phi_{01}(t_1), \Phi_{10}(t_2)]
\end{equation}
By naming $C=\int_{-\infty}^{+\infty} dt \: \Phi(t)$ and $C_{00}=\int_{-\infty}^{+\infty} dt \: \Phi_{00}(t)$, we introduce
\begin{equation}\label{eq:contraction1}
\widetilde{K}_T = {1\over 2} (C-C_{00})^2 + {1\over 2} \int_{-\infty}^{+\infty} \!\!\! dt_1 \int_{-\infty}^{t_1} \!\!\! dt_2\;  [(\Phi-\Phi_{00})(t_1), (\Phi-\Phi_{00})(t_2)]
\end{equation}
where $(\Phi-\Phi_{00})(t)=\Phi(t)-\Phi_{00}(t)$. It follows easily that
\begin{equation}
K_T=P_0\widetilde{K}_T P_0.
\end{equation}
Now again through the Trotter product formula, the theorem would follow from the fact that $\widetilde{K}_T$ generates a contraction semigroup on $\mathcal{B}$. Indeed,
\begin{eqnarray}
P_0 e^{K_T t} &=& \lim_{n\rightarrow\infty} P_0 \{1+ P_0 \widetilde{K}_T P_0 t/n +O(n^{-2})\}^n \nonumber\\
&=& \lim_{n\rightarrow\infty} \{P_0 e^{\widetilde{K}_T t/n} P_0\}^n,
\end{eqnarray}
so that, since $\|P_0\|= 1$, we obtain
\begin{equation}
\|P_0 e^{K_T t}\|\leq \| e^{\widetilde{K}_T t} \|
\end{equation}
for all $t>0$. So we shall prove that $\widetilde{K}_T$ generates a contraction semigroup on $\mathcal{B}$ by showing that each of the two terms in (\ref{eq:contraction1}) does.

In order to do that, we note that both $C$ and $C_{00}$ (and thus also $C-C_{00}$) generate a one-parameter group of isometries on $\mathcal{B}$, as can be easily seen through repeated use of the Trotter formula for both positive and negative $t$, recalling we assume $A$ and $A_{00}$ generate isometries, and $U_t$ are isometries. In fact, recalling $C$ is bounded (as $C_{00}$), and that $\Phi(t)\rightarrow 0$ for $t\rightarrow\pm\infty$, with obvious notation one has
\begin{equation}
e^{C t}=\lim_{\sup\{\Delta_i\}\rightarrow 0} \lim_{n\rightarrow\infty} \{\prod_j e^{\Delta_j \Phi(t_j) t/n} \}^n
\end{equation}
and for all subdivisions $\{\Delta_j\}$ and natural $n$
\begin{equation}
\|\{\prod_j e^{\Delta_j \Phi(t_j) t/n} \}^n\|\leq \prod_j \|e^{\Delta_j \Phi(t_j) t/n} \|^n\leq 1,
\end{equation}
as $e^{\Phi(t') t}$ is a contraction on $\mathcal{B}$ for all real $t'$ and $t$. This in turn follows because $\Phi(t')$ satisfies all the hypotheses of the Hille-Yosida theorem~\cite{bratteli_robinson} as, being bounded for every finite $T>0$, it is in particular weakly closed and densely defined, and for all real $\alpha\geq\|A\|\geq\|\Phi(t')\|$ the range $R(\beta 1- \Phi(t'))$ of $\beta 1- \Phi(t')$ is the entire space $\mathcal{B}$ (see propositions 3.1.1 and 3.1.6 in~\cite{bratteli_robinson}). Moreover, for all real $\alpha$ and $b\in\mathcal{B}$ we have
\begin{equation}
\|(1-\alpha \Phi(t'))b\|\geq \|(1-{\alpha\over\sqrt{\sqrt{\pi}T}} A)U_t b\| \geq \|b\|
\end{equation}
because $A$ generates isometries and $U_{\pm t}$ are isometries. Then the Hille-Yosida Theorem asserts that $\exp\{\Phi(t') t\}$ is a one-parameter contraction semigroup (in the $t$ variable) for every real $t'$. Since the inequalities above hold for all real $\alpha$, Corollary 3.1.19 of~\cite{bratteli_robinson} further implies that $\exp\{\Phi(t') t\}$ (and hence also $\exp\{C t\}$) is a one-parameter group of isometries on $\mathcal{B}$. The same arguments imply that $C_{00}$ generates a one-parameter group of isometries on $\mathcal{B}$.

Now the first term in (\ref{eq:contraction1}) generates a weakly-continuous (and hence strongly continuous) semigroup of contractions through the Hille-Yosida Theorem~\cite{bratteli_robinson} as, being bounded for every finite $T>0$, it is weakly closed and densely defined, and for all real $\alpha>\|\widetilde{K}_T\|$ the range $R(\alpha 1- (C-C_{00})^2)$ of $\alpha 1- (C-C_{00})^2$ is the entire space $\mathcal{B}$ (see proposition 3.1.6 in~\cite{bratteli_robinson}). Moreover, by writing $1-\alpha(C-C_{00})^2=(1-\sqrt{\alpha}(C-C_{00}))(1+\sqrt{\alpha}(C-C_{00}))$ for $\alpha>0$, and using the fact that $C-C_{00}$ generates isometries, we see that
\begin{equation}
\|(1-\alpha(C-C_{00})^2)b\|\geq \|(1+\sqrt{\alpha}(C-C_{00}))b\| \geq \|b\|
\end{equation}
for any $\alpha>0$ and $b\in\mathcal{B}$.

To treat the remaining term in (\ref{eq:contraction1}), we denote $D_t=(\Phi-\Phi_{00})(t)$, so that by the Hille-Yosida Theorem $D_{t'}$ generates a one-parameter group of isometries for any real $t'$. Now, according to Lemma \ref{lemma:contraction}, $[D_{t_1},D_{t_2}]$ generates a one-parameter group of isometries for every $t_1$ and $t_2$. The integrals on such $t_1$ and $t_2$ can again easily be shown, as before, to generate isometries by the Trotter formula (they generate contractions for all $t\in\mathbb{R}$).

This completes the proof.
$\quad\Box$ \\

\subsection{Comments to the Theorem}
\begin{itemize}
\item It is evident that none of $K_{(\alpha,q,T)}$ generates contractions, so this theorem actually establishes a "uniqueness result" in the class of all the possible semigroup approximation of the exact projected dynamics (\ref{eq:exact}). Moreover, having a contraction semigroup at disposal is very important for physical applications, where one is willing to study the limit dynamics at all times, including steady states. It is well known indeed that a Quantum Dynamical Semigroup in a von Neumann algebra~\cite{lindblad} is, in particular, a contraction semigroup.
\item We have shown that our dynamical \emph{time} coarse-graining procedure does lead to a \emph{dissipative} generator for the irreversible dynamics. Physically one could say that a certain scale is fixed (the transition time), in which the subsystem is not able to distinguish the details of the evolution. Performing a time average on that scale allows one to eliminate the exponentially growing "off-diagonal terms" in the generator, thus obtaining a genuine dissipative process. Probably this could be explained by saying that $K_{(\alpha,q,T)}$ "sees" pairs of transitions $A_{01}$ and $A_{10}$ instantaneously, at two specific and different instants inside the temporal transition window/scale, resulting in a subsystem internal transition which is strongly asymmetric (and singular) in time. Through the dynamical time average $K_{T(\lambda)}$ instead, subsystem internal transitions take place smoothly and homogeneously in time.
\item Given that $K_T=P_0\widetilde{K}_T P_0$ with $\widetilde{K}_T$ in (\ref{eq:contraction1}) is the sum of two terms, the first term is dissipative, whereas the other one is a second order energy renormalisation, generating a one-parameter group of isometries.
\end{itemize}

\section{Summary and Conclusion}
We have considered the class of quantum mechanical master equations for Physical Subsystems in the weak coupling limit.
In the first part we have taken this to be a perturbed one-parameter group of isometries projected on a Banach subspace.
In that limit, we have shown that the memory terms of the integral equation for the evolution operator can be approximated in a variety of different ways: the resulting class of one-parameter semigroups includes the previous literature.
Then through the introduction of a dynamical time averaging map we have found explicitly a generator that $i)$ is able to correctly approximate the exact projected dynamics in the weak-coupling limit, $ii)$ is always well defined, irrespective of the subsystem spectral properties and dimensions, $iii)$ accounts also for first order contributions, $iv$ boils down to previous literature results in case of discrete spectrum, and more importantly $v)$ gives rise to a contraction semigroup on the projected Banach subspace.

Our results are of very general nature, and open the way to the study of a variety of different extensions (beyond second order, time dependent free dynamics, to name a few) and physical applications in the field of operator algebras, also greatly linked to nowadays technologies.

\section*{Acknowledgments}
We wish to thank Prof. Fausto Rossi (NTL, Phys. Dept., Politecnic of Turin) for profound and enlightening discussions on all the important concepts in this work. We would like to thank Prof. Hisao Fujita Yashima (Dept.
Mathematics, University of Turin) for having offered so many days of invaluable help and discussions to the author. We furthermore wish to thank Dr. Taj Mohammad for providing the figures.

\end{document}